%% file: main.tex
\documentclass[letterpaper,twocolumn,10pt]{article}
\usepackage{usenix}
\usepackage{amsmath}
\usepackage{booktabs}
\usepackage{multirow}
\usepackage{array}
\usepackage{caption}
\usepackage{graphicx}
\usepackage{subcaption}
\usepackage{makecell}
\usepackage{threeparttable}
\usepackage{tabularx}
\usepackage{amssymb}
\usepackage{xcolor}
\usepackage[capitalize]{cleveref}

\makeatletter
\AtBeginDocument{%
  \@ifundefined{cref@old@label@in@display}{}{%
    \def\label@in@display@noarg#1{\cref@old@label@in@display{#1}}%
  }%
}
\makeatother

\usepackage[available,functional]{usenixbadges}

\newcommand{\alias}{\texttt{InjectEave}}

\definecolor{revcol}{rgb}{0,0,0}
\newcommand{\rev}[1]{\textcolor{revcol}{#1}}

\begin{document}

\date{}

\title{\Large \bf Injected and Leaked: Actively Inducing Side-Channel Leakage Using Electromagnetic Injection and Hardware Nonlinearity}

\author{
{\rm Haoran Yan}$^{1}$, {\rm Ziyu Shao}$^{1}$, {\rm Shuhao Zhang}$^{1}$, {\rm Qinhong Jiang}$^{2\dagger}$, {\rm Yan Long}$^{1\dagger}$\\[0.4ex]
$^{1}$The Hong Kong University of Science and Technology (Guangzhou) \\ $^{2}$The Hong Kong Polytechnic University\\[0.4ex]
\{hyan097, szhang515\}@connect.hkust-gz.edu.cn; \quad ziyushao@hkust-gz.edu.cn \\
 qinhong.jiang@polyu.edu.hk;\quad yanlong@hkust-gz.edu.cn
}

\maketitle

\begingroup
  \renewcommand{\thefootnote}{} 
  \footnotetext{\hspace{-1.5em}$^{\dagger}$ Corresponding authors.} 
\endgroup

\begin{textblock}{3.2}(0.75,10.05)
  \noindent\footnotesize\itshape
  Original publication: USENIX Security '26,\\[-0.15ex]
  USENIX Association, 2026.
\end{textblock}

\input{sections/0-abstract}

\input{sections/1-Introduction}

\input{sections/2-Background}

\input{sections/3-injection-induced-em-side-channel-leakage}

\input{sections/4-eavesdropping-design}

\input{sections/5-evaluation-in-a-laboratory-setting}

\input{sections/6-audio-eavesdropping-in-the-wild}

\input{sections/7-Discussion}

\input{sections/8-Conclusion}

\input{sections/9-ethical-considerations}

\bibliographystyle{plain}
\bibliography{refs}

\end{document}

%% file: sections/0-abstract.tex
\begin{abstract}
Electromagnetic (EM) side-channel leakage and injection are typically treated as distinct physical phenomena, threatening data confidentiality and integrity respectively. This work investigates how EM injection can be used to amplify side-channel leakage that is otherwise infeasible. We introduce a novel framework for \textit{Injection-Induced EM Side Channels} to enable integrated, closed-loop EM security analysis. Our theoretical modeling and experimental measurements reveal that nonlinear hardware components, such as ubiquitous amplifiers, analog-to-digital converters, and power converters, can modulate secret electrical signals onto an injected EM carrier and thus upconvert low-frequency secrets into measurable EM emissions. By tuning the injection frequency and amplitude, adversaries gain the ability to actively shape the effective spectrum and entropy of the resulting leakage. We design \alias{} attack and demonstrate eavesdropping on the audio played through wired and wireless headphones from up to 30~m away with accessible RF equipment, as well as in through-wall scenarios, and characterize injection-induced EM leakage of other low-frequency secrets such as power consumption of smart home devices and analog sensor inputs. Case studies further demonstrate how the proposed techniques enable closed-loop eavesdropping and manipulation of landline-phone conversations. Finally, we analyze the broader security challenges and mitigations.
\end{abstract}

%% file: sections/1-Introduction.tex
\section{Introduction}
This work investigates the new problem statement of exploiting active electromagnetic (EM) injection and the inherent nonlinearity of computer hardware to reshape the capability of conventional electromagnetic side-channel analysis. Side-channel analysis has become one of the most important types of security analysis methodologies, exploiting the non-ideal abstractions of computer systems to compromise confidentiality and achieve unauthorized access to data internal to a protected device~\cite{agrawal2002side,standaert2009introduction}. By collecting and analyzing signals unintentionally produced by the physical operations of computer hardware, such as sound~\cite{al2016acoustic, backes2010acoustic,bolton2023characterizing,long2023side}, light~\cite{ferrigno2008aes,loughry2002information,long2023private}, and electromagnetic emissions~\cite{kuhn2002optical, long2024eye, vuagnoux2009compromising}, adversaries would be able to infer critical information of cryptographic operations~\cite{gnad2019leaky,camurati2018screaming,genkin2015stealing}, confidential input data~\cite{jin2021periscope, halevi2015keyboard, bolton2023characterizing}, and private user identity~\cite{long2024eye,li2025emiris,ni2023recovering, li2023magfingerprint}.

\begin{figure}
    \centering
    \includegraphics[width=1\linewidth]{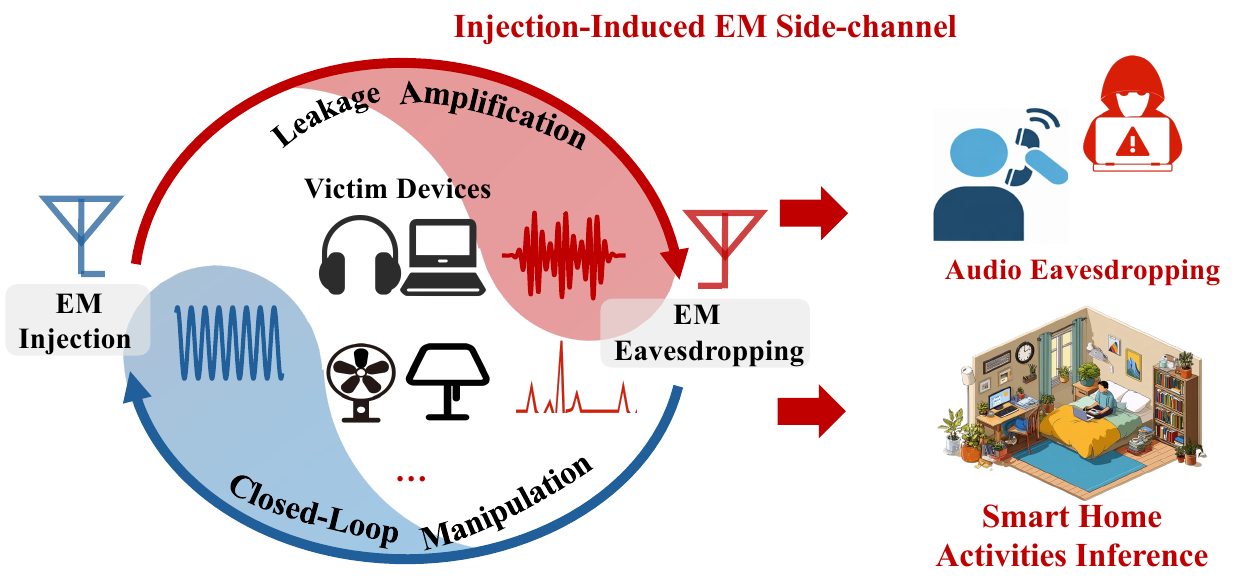}
    \caption{Injection-induced EM Side Channel utilizes active EM injection and ubiquitous hardware nonlinearity to produce controllable EM leakage, enabling unconventional EM eavesdropping vectors against low-frequency analog secrets.}
    \label{fig:overview}
\end{figure}

Among various methods of side channels, EM side channels present a highly pervasive and impactful attack surface, as all modern computer systems rely on current and voltage variations in electrical circuits to perform all computations. The resulting time-varying EM fields inevitably radiate into the surrounding environment and propagate through the air to nearby adversaries. For example, the security community has shown the feasibility of exploiting EM side-channel leakage to eavesdrop on a wide range of secret information, such as screen displays~\cite{kuhn2005electromagnetic,liu2020screen}, keyboard and touchscreen interactions~\cite{vuagnoux2009compromising, jin2021periscope}, users' biometrics~\cite{li2025emiris,ni2023recovering,xu2025empalm}, and even confidential video streams of smart home cameras~\cite{long2024eye}.  

Despite the massive theoretical attack surface, existing EM side-channel research has revealed a critical limitation in the range of applicable eavesdropping distances and observable types of information, {\color{revcol} especially on secrets in the form of low-frequency analog signals}. Specifically, the EM energy that can propagate to external adversaries is solely determined by the target's internal characteristics, including the amplitude of the current/voltage that carries the secret information, the frequency of the internal electrical signals, and the EM transfer efficiency of the target's circuits that act as unintentional radiating antennas. This major limitation is rooted in the threat model assumption that the side-channel analyzer can only passively observe the EM leakage of a target device. 
As a result, \textit{conventional {\color{revcol} passive} side-channel eavesdroppers face the seemingly ``unsolvable'' problem of low signal-to-noise ratio (SNR)}, treating better EM receivers as their only measure for improving EM side-channel capabilities. 

Toward overcoming this challenge, this work rethinks the passive eavesdropping paradigm and provides a new analytical framework that employs active EM injection to induce and amplify EM side-channel leakage {\color{revcol} of analog electrical signals} in controllable ways. The key insight behind our approach is that a fundamental frequency mismatch between internal secret signals and the circuit's efficient EM coupling bands creates a physical barrier, significantly limiting the secret energy that can \textit{\textbf{leave}} the device. Meanwhile, we observe that existing injection research~\cite{kune2013ghost,tu2018injected, jiang2024ghosttype} has shown how external EM signals can be designed to be at the most efficient coupling frequencies to \textbf{\textit{enter}} the target device. Importantly, injected EM signals could be unintentionally demodulated by nonlinear hardware such as amplifiers, allowing adversaries to use EM carriers to inject false \textbf{\textit{low-frequency analog signals}} into target systems.
{\color{revcol} This is related to the recent concept of active side-channel analysis, where adversaries generate signals to illuminate a device and analyze the reflected response to recover digital data such as serial bits~\cite{kaji2023echo, pu2025your} or cryptographic information~\cite{monfared2023leakyohm, kitazawa2026active}. However, such a reflection-based impedance-change model is limited to coarse-grained binary impedance-state analysis and cannot characterize waveform-level leakage of continuous analog secrets.

Building upon these works and the significant existing gaps, we formulate \textit{Injection-Induced EM Side Channel}. Unlike reflection-based approaches, our method characterizes how internal electrical signals are modulated onto an injected carrier through hardware nonlinearity and subsequently converted into secret-bearing EM leakage. This bridges the theoretical and experimental gaps between side-channel leakage and EM injection, establishing the physical basis for injection-induced eavesdropping, even for ubiquitous analog signals.} We hypothesize that \textit{nonlinear computer hardware can modulate {\color{revcol} analog} secret information, in the form of electrical inputs of these hardware components, onto injected carriers, which could then emit and propagate back to side-channel eavesdroppers.}

If this hypothesis were true, then EM side-channel eavesdroppers would be able to actively control their injected EM carriers to break through the target devices' EM security boundaries. We characterize this threat model with experiments on four types of the most typical nonlinear hardware components found in computer systems, including amplifiers, analog-to-digital converters, power converters, and switching MOSFETs. Our measurements verify that injection-induced side channels enable unconventional attack vectors, such as eavesdropping on secret information with EM frequencies on the order of 10~Hz--10~kHz, which itself could be too low to propagate to external eavesdroppers. 
Our theoretical modeling further provides a framework for analyzing threats against the most common analog data interfaces, such as audio output and input, control signals of actuators, and even device power traces (\cref{fig:overview}). 

Despite these new capabilities, our tests show that signals eavesdropped with this approach unavoidably suffer from higher-order inter-modulation that adds harmonics of an original signal to its spectrum. This nonlinearity-specific distortion poses unique challenges to eavesdroppers who aim to recover wideband signals, such as human speech audio. We design \alias{} to recover higher-fidelity secrets. \alias{} utilizes a diffusion-based denoising model and training data simulated by our developed quantitative model. Tests on physically collected speech audio data show notable improvements in several audio quality metrics. 

Our evaluation in lab settings first identifies typical low-frequency analog secrets carried by 11 commercial off-the-shelf (COTS) household devices. Audio of headphones and landline phones could leak both speaker identity and spoken content. Control signals and power consumption traces of IoT devices, such as smart fans and lamps, could leak personal activities in households. Our tests show that injection-induced side-channel attacks could eavesdrop on the majority of these devices from over 2~m away and through walls, with a maximum distance of 30~m for recovering intelligible headphone audio. 
Our case studies of audio eavesdropping in several personal, public, and work scenarios further demonstrate its security consequences in the wild\footnote{Demos and data are available at: \href{https://injecteave.github.io/}{https://injecteave.github.io/}}. An example of landline phones showcases the new capability of integrating EM injection and injection-induced leakage to achieve closed-loop, context-aware eavesdropping and manipulation on private conversations. 
Finally, we discuss the other possible analog and digital secrets that need to be further threat-modeled under this emerging threat of injection-induced EM side channel, and analyze the possible mitigations, highlighting the urgent need for systematically examining the pervasive threat of injection-induced EM side channels. The main contributions of this work are summarized as follows:

\begin{itemize}
    \item Theoretical framework for \textbf{\textit{Injection-Induced EM Side Channels}} of analog secrets. We identify ubiquitous nonlinear hardware as the root cause of this new phenomenon, enabling future research to integrate closed-loop data eavesdropping and manipulation analysis.
    \item Technical design and implementation of \alias{}. Our design exemplifies how to exploit this new vulnerability, enabling through-wall eavesdropping on low-frequency analog secrets such as speech audio and activities of smart home devices, and achieving a long-range audio eavesdropping capability of up to 30~m.

    \item Characterization of threats and mitigation. Our evaluation on 11 COTS devices gauges this new threat model's impact, based on which we analyze possible defense methodologies for stronger EM security protection. 
\end{itemize}

%% file: sections/2-Background.tex
\section{Background}

This section introduces the motivation for exploring the new techniques of injection-induced side channels, by reflecting on the history and limitations of conventional side channels and the new opportunities. 

\subsection{Conventional EM Side Channel}

A conventional EM side channel is an unintentional one-way communication channel between the target device and an eavesdropper. Operations of computer systems are physically implemented by changing voltages (and equivalently, currents) in hardware circuits, which generate varying electromagnetic fields. Then, an electrical trace, such as a wire on the PCB or a communication cable, can act as an unintentional antenna that unwittingly sends the internal EM energy to the surrounding environment. Denoting the internal voltage signal of the secret as $V_{sec}(t)$, the side-channel leakage process can be represented as: $V_{eav}(t) = h_{tx}(V_{sec}(t))$, where $h_{tx}(t)$, with its frequency-domain representation denoted as $H_{tx}(f)$, is the transfer function describing the frequency response of the unintentional leakage source and the EM propagation path. 

\begin{figure}[t!]
    \centering
    \includegraphics[width=0.99\linewidth]{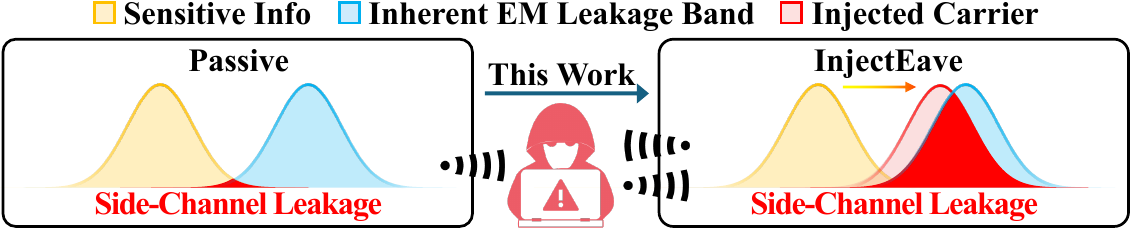}
    \caption{Injected EM carriers can piggyback secret signals, 
    overcoming the mismatch between target signal frequencies and efficient EM emissions.}
    \label{fig:em-comparison}
\end{figure}

While adversaries trying to get higher-amplitude signals have control over $H_{tx}$ to some degree by reducing their physical distance from the target, or employing higher-gain receiving antennas, the majority of the usable leakage signal is determined by the frequency and amplitude of $V_{sec}(t)$ itself, and the efficiency of $H_{tx}$ over the frequency bands of $V_{sec}(t)$. \textit{For eavesdroppers, unfortunately, the efficient frequency of $H_{tx}$ and the frequency of the interested signals $V_{sec}(t)$ often face a mismatch}, as shown in~\cref{fig:em-comparison}. Taking human speech audio signals as an example, the target signals are in the range of 20~Hz--20~kHz, which is far from the feasible EM frequency range (at least on the order of MHz) of most unintentional antenna structures within the target device. Furthermore, the feasible eavesdropping distances have been limited due to the increasingly lower operation voltages of low-power miniaturized electronics~\cite{chandrakasan1992low} that reduced the amplitude of $V_{sec}(t)$, and stronger EM shielding~\cite{ott2011electromagnetic} that reduced the amplitude of $H_{tx}$ in newer computer systems.

As a result, EM side-channel leakage has so far remained a notable risk mostly for high-voltage digital data transmissions, such as computer display images~\cite{kuhn2005electromagnetic} and keyboard inputs transmitted over USB cables~\cite{vuagnoux2009compromising}, revealing a gap in side-channel analysis capability for low-frequency analog secrets.

\subsection{EM Injection and Device Nonlinearity}
EM injection is a technique used to physically inject false analog signals into computer hardware. In 2013, Foo Kune et al.~\cite{kune2013ghost} demonstrated that amplitude-modulated EM waveforms can accurately change the analog sensor readings of implanted defibrillators and microphones. The core of EM injection methodologies is the exploitation of nonlinear hardware components for addressing the mismatch between the frequency of intended malicious signals and the effective EM injection frequency bands, which are, again, determined by the target hardware's EM frequency response. 

For example, \cite{kune2013ghost} was able to inject kHz-range fake speech audio into microphone readings, where audio signals are amplitude-modulated onto EM carriers of 840~MHz. The target device's electrical traces act as unintentional receiving antennas that pick up the modulated carriers. When the received high-frequency signals pass through microphones' amplifiers, which have unmodeled nonlinear input-output relationships, the baseband false audio signals will be demodulated to the original frequency range and become inputs of microphones and other sensors.

Since then, EM injection has been widely considered as a means for compromising data availability and integrity, such as injecting false keystrokes~\cite{jiang2024ghosttype,zhang2024virtual}, inducing fake touchscreen inputs~\cite{shan2022invisible,maruyama2019tap}, and altering camera images~\cite{jiang2023glitchhiker, zhuang2025rfeyed}. However, our work discovers and characterizes the hidden capability of EM injection for inducing side-channel leakage and enhancing security analysis on data confidentiality: \textbf{\textit{the nonlinearity of hardware may not only demodulate information from EM carriers, but also modulate secrets onto EM carriers}}. This new perspective addresses exactly the knowledge gap in the need for more effective EM side-channel methodologies that can capture low-frequency analog secrets.

%% file: sections/3-injection-induced-em-side-channel-leakage.tex
\section{Injection-Induced EM Side Channel}
\label{sec:mechanism}

Based on the observations and analysis above, this section provides the threat model and formulation of injection-induced EM side channels, and characterizes its feasibility in widely found nonlinear hardware within computer systems. 

\begin{figure}[t]
    \centering
    \includegraphics[width=\linewidth]{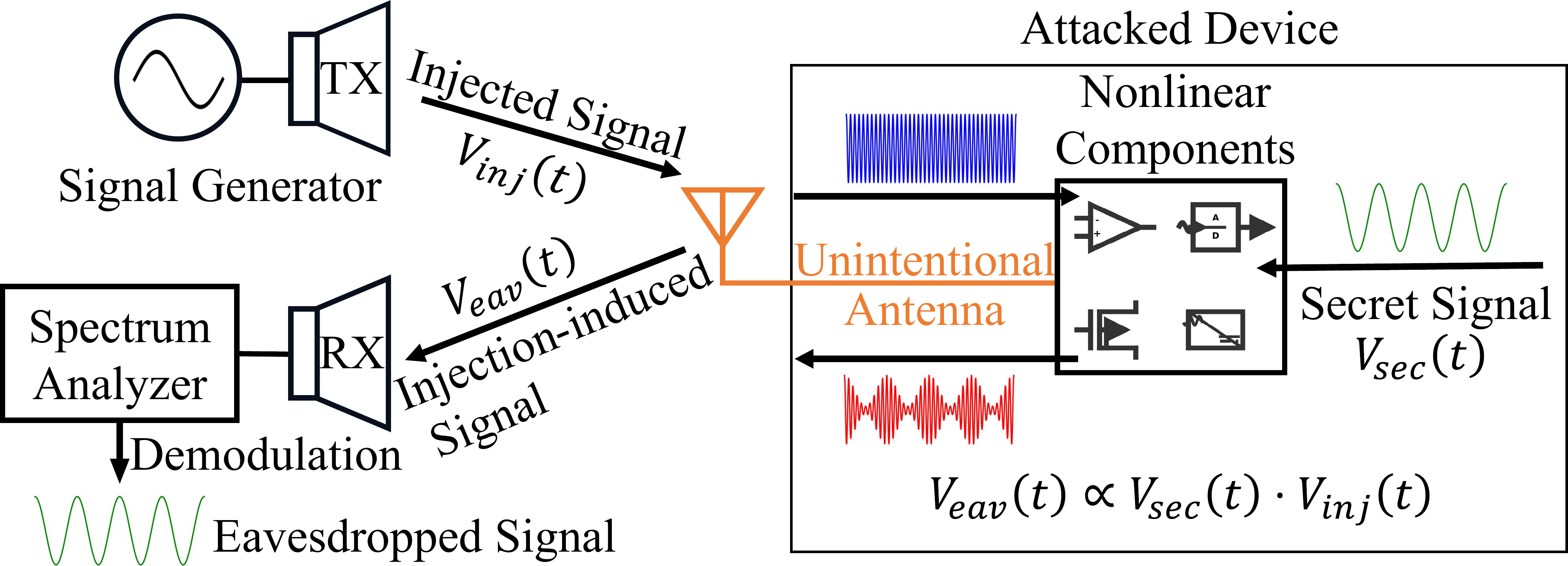}
    \caption{Model of the injection-induced leakage process.}
    \label{fig:leakage_model}
\end{figure}
\subsection{Threat Model}

We hypothesize that EM signals injected into nonlinear computer hardware can mix with and thus piggyback secret information within the target system; the modulated EM signals will then leak to side-channel adversaries. In particular, the injected EM energy boosts the amount of side-channel leakage, enabling adversaries to get information previously inaccessible with conventional EM side-channel methods.

\textbf{Adversary's Objective.} The objective of the adversary is the same as that in conventional EM side-channel analysis~\cite{kuhn2005electromagnetic, vuagnoux2009compromising, long2024eye}: inferring confidential information about a computer system's operations by analyzing the EM signals the adversary can collect. Our analysis in this work focuses on secrets in the form of analog signals, particularly the low-frequency signals such as human speech audio (below 20~kHz), power consumption and actuation control signals (below 200~Hz), which are known to be highly challenging targets due to the spectral mismatch between these secrets' frequencies and efficient EM leakage frequencies (MHz or GHz range)~\cite{wheeler2006fundamental, kuhn1998soft}. 

\textbf{Adversary's Capability.} We assume the adversary has a set of readily available commercial equipment that is able to both send EM injection signals and receive modulated EM emissions.
The added EM injection capability is the only difference from the assumed capability of conventional EM side-channel adversaries. The equipment often includes antennas, RF sources such as software-defined radio devices (e.g., USRP~\cite{ettus2015universal}), and potentially more advanced spectrum analyzers and signal generators commonly found in RF research labs. As in conventional EM side-channel research, we assume the adversary has prior knowledge of the target device's model and can acquire a similar device for profiling its effective EM injection and emission frequencies.

\subsection{Leakage Modeling}
\label{sec:leakage_model}

We formalize the Injection-Induced Side Channel through an \textit{Injection-Modulation-Emission} model, as depicted in~\cref{fig:leakage_model}. 

\textbf{Injection Coupling.} The adversary generates an injection carrier $V_{inj}(t)$ at frequency $f_c$: $V_{inj}(t) = A_{inj} \cos(2\pi f_c t)$. 
The target device's electrical traces act as unintentional receiving antennas, where the injection coupling efficiency is governed by a frequency-dependent injection transfer function $H_{rx}(f)$ (equivalently, $h_{rx}(t)$ in the time domain). The induced voltage $V_{c}(t)$ at the input of the vulnerable nonlinear component is:
\begin{equation}
    V_{c}(t) = |H_{rx}(f_c)| \cdot A_{inj} \cos(2\pi f_c t + \phi)
\end{equation}
Consequently, the total signal $V_{in}(t)$ present at the input terminal of the nonlinear component is the superposition of the original secret signal and the coupled carrier:
\begin{equation}\label{eq:Vin}
    V_{in}(t) = V_{sec}(t) + V_{c}(t)
\end{equation}

\begin{figure}[t]
    \centering
    \includegraphics[width=\linewidth]{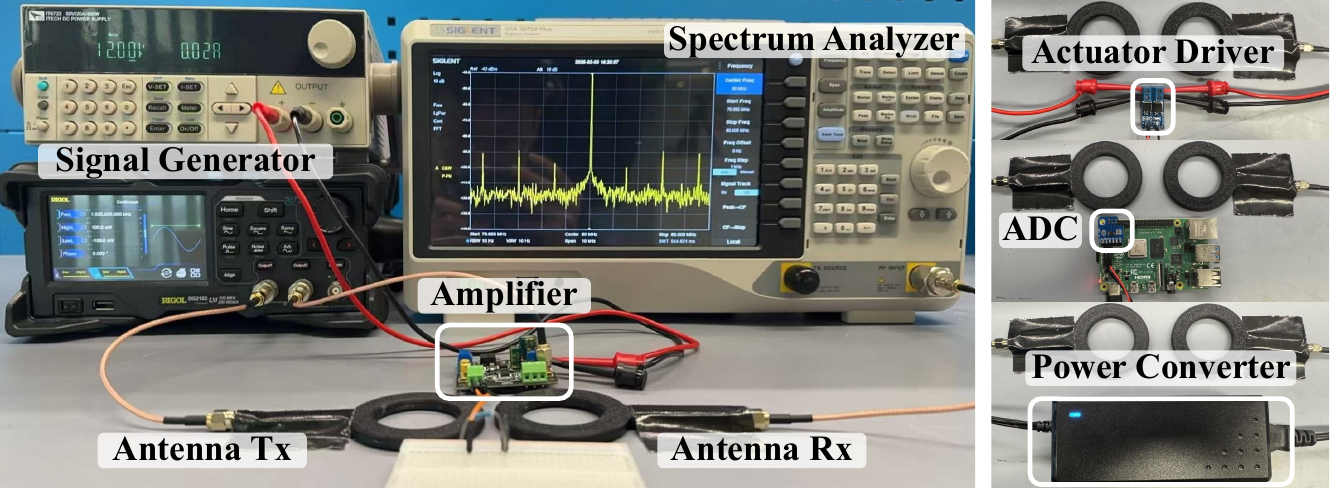}
    \caption{Feasibility tests on the most common and ubiquitous nonlinear hardware components.}
    \label{fig:feasibility test setup}
\end{figure}

\begin{figure*}[th!]
    \centering
    \includegraphics[width=1.0\linewidth]{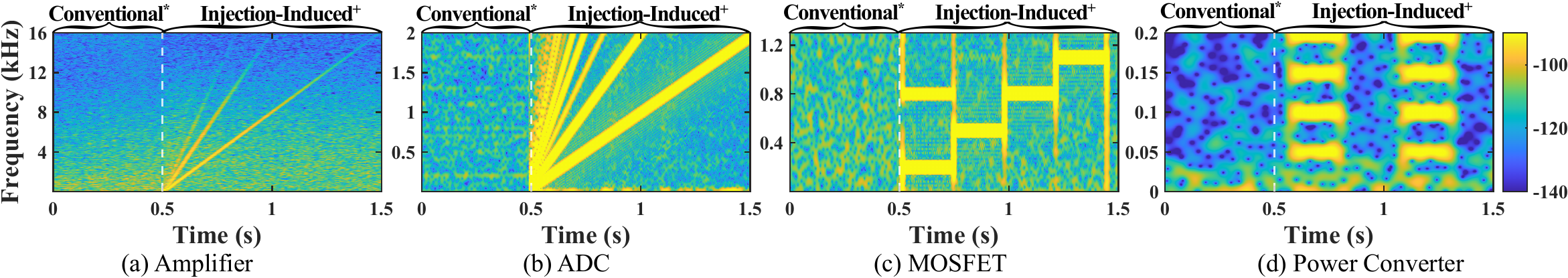}
    \caption{{\color{revcol}Ubiquitous nonlinearities in commodity computing hardware can unintentionally leak secret analog information under EM injection. $^{*}$Directly measured baseband signal. $^{+}$Baseband signal recovered by down-converting from the carrier frequency. The two segments of signals are concatenated together post-hoc for easier comparison.} }
    \label{fig:feasibility test.}
\end{figure*}

\textbf{Nonlinear Modulation.}
Following the series expansion model for semiconductor nonlinearity established in prior EMI research~\cite{kune2013ghost}, we approximate the transfer function of the nonlinear hardware component as:
\begin{equation} \label{eq:serial}
    V_{out}(t) = \sum_{k=0}^{\infty} \alpha_k V_{in}^k(t) = \alpha_0 + \alpha_1 V_{in}(t) + \alpha_2 V_{in}^2(t) + \dots
\end{equation}
where $\alpha_k$ represents the $k$-th order coefficient. While the linear term $\alpha_1$ represents intended signal conditioning such as amplification, the quadratic term $\alpha_2$ and higher-order terms induce inter-modulation. Taking the quadratic term for example, substituting~\cref{eq:Vin} into the quadratic term yields:
\begin{equation}
\label{eq:expansion}
\begin{split}
    \alpha_2 V_{in}^2(t) &= \alpha_2 [V_{sec}(t) + V_{c}(t)]^2 \\
    &= \alpha_2 [V_{sec}^2(t) + V_{c}^2(t) + \underbrace{2 V_{sec}(t) V_{c}(t)}_{\text{AM Modulation}}]
\end{split}
\end{equation}
The cross-product term in~\cref{eq:expansion} exemplifies how the secret could be modulated onto the injected carrier. Denoting the aggregate signals carrying the modulated secret as $V^{c}_{sec}(t)$, the model shows: 
\begin{equation} \label{eq:Vmod}
    V^{c}_{sec}(t) = 2 \alpha_2 V_{sec}(t)V_{c}(t) + V^{c}_{hi}(t),
\end{equation}
where $V^{c}_{hi}(t)$ represents the high-order inter-modulation products associated with $\alpha_3, \alpha_4$, etc. 
This process effectively up-converts the spectral energy of $V_{sec}(t)$ from the baseband to the sidebands centered at $f_c$. 

\textbf{Carrier Emission.}
The modulated signal $V^{c}_{sec}(t)$ propagates through the device's conductive paths and emits at other electrical interconnects, which act as unintentional transmitting antennas. The leakage efficiency is determined by the emission transfer function $H_{tx}(f)$. The final leakage signal $V_{eav}(t)$ observed by the adversary is:
\begin{equation} \label{eq:Veav}
    V_{eav}(t) = 2 \alpha_2\cdot h_{tx}( V_{sec}(t)V_{c}(t)) + h_{tx}(V^{c}_{hi}(t))
\end{equation}
When only considering the dominant second-order inter-modulation for simplicity, the leakage amplitude $|V_{eav}(t)|$ can be expressed as a function of the system parameters:
\begin{equation} \label{eq:effi}
   |V_{eav}(t)| \propto \underbrace{|H_{tx}(f_c) H_{rx}(f_c)|}_{\text{Coupling Efficiency}} \cdot \underbrace{|\alpha_2|}_{\text{nonlinearity}} \cdot \underbrace{|V_{sec}(t)|\cdot|V_{inj}(t)|}_{\text{Signal Amplitude}}
\end{equation}

The modeling reveals that even if the original secret signals' amplitudes are low and their frequencies are significantly lower than the efficient emission frequencies of $H_{tx}(f)$, the adversary can amplify the leakage by tuning the injection frequency $f_c$ to maximize the compound efficiency product $|H_{tx}(f_c) H_{rx}(f_c)|$, as well as by increasing the amplitude $A_{inj}$ of EM injection $V_{inj}(t)$. 

\subsection{Feasibility Analysis}~\label{sec:feasibility}
To verify this hypothesis of injection-induced leakage caused by inter-modulations of $V_{sec}(t)$ and $V_{inj}(t)$, we individually characterized the leakage behavior of four types of the most common nonlinear hardware found in computer systems, including (1) amplifiers, (2) analog-to-digital converters (ADCs), (3) switching Metal-Oxide-Semiconductor Field-Effect Transistors (MOSFETs), and (4) power converters. As shown by the setup in~\cref{fig:feasibility test setup}, we used two near-field electromagnetic probes to inject EM energy into and receive emissions from the nonlinear components. 

\subsubsection{Susceptibility of Common Nonlinear Electronics}
\label{sec:nonlinear_electronics}

For each nonlinear component, we measured the electromagnetic emissions under both conventional passive eavesdropping conditions and our proposed injection-induced conditions. This comparative analysis aims to illustrate the leakage-enabling capability provided by the injection-induced side channels. {\color{revcol} We first perform wideband RF spectrum sweeps from 0 to 2~GHz across all four components under passive conditions. No discernible leakage correlated with the target secret signals is observed, indicating that low-frequency secret signals are inherently difficult to recover through passive EM emissions alone.} In addition, we also measured and observed no leakage signals when the nonlinear components were replaced by linear resistance loads, confirming that hardware nonlinearity is the key for injection-induced leakage.

\textbf{(1) Amplifier.} Amplifiers are well-known nonlinear devices~\cite{kune2013ghost, tu2019trick} often found in data interfaces such as audio output and sensor input circuits. 
We selected Analog Devices' AD623, a widely used rail-to-rail instrumentation amplifier, as our primary target. To characterize its nonlinear response in a controlled setting, we used a signal generator to directly input a baseband frequency sweep signal ($V_{sec}(t)$) into the amplifier's input terminal. The sweep secret signal ranged from 0 to 16~kHz with an amplitude of 200~mV. Simultaneously, we targeted the device with an EM injection carrier $V_{inj}(t)$ at 80~MHz. 
The results, visualized in~\cref{fig:feasibility test.}~(a), clearly show the spectral content of the original secret signal, together with distinct harmonic components ($2\times f_{secret}$, $3\times f_{secret}$, ...).
This observation further provides proof of the hardware's nonlinearity. In contrast, conventional eavesdropping of the 0--16~kHz analog secret shows no visible signals on the receiver.

\textbf{(2) ADC.} Analog-to-digital converters, another type of ubiquitous component in modern digital computer systems that process inputs of analog physical information, also have nonlinear characteristics.  
We used Texas Instruments' ADS1115, a common 16-bit ADC, as the test target. Similar to the amplifier test, we input a baseband sweep signal $V_{sec}(t)$ while targeting the device with the RF carrier at 80~MHz. As shown in~\cref{fig:feasibility test.}, clear secrets and their harmonic signals are observed in the injection-induced leakage, while the conventional side channel analysis receives no useful information. Notably, the results in~\cref{fig:feasibility test.}~(d) reveal that leakage persists even when the injection bandwidth far exceeds the ADC's sample rate, which is 860~Hz, revealing that modulation occurs in the continuous-time analog front-end.

\textbf{(3) Switching MOSFET in Actuators.} Driver circuits of actuators allow computer systems to control external hardware, such as motors in IoT devices. These circuits typically employ nonlinear power MOSFETs acting as high-speed switches to provide control signals, such as pulse width modulation (PWM). To examine their injection-induced leakage, we constructed a representative driver circuit using a discrete MOSFET driving a resistive load. An Arduino generates the baseband secret signal, $V_{sec}(t)$, in the form of a frequency-stepped square wave (shifting between 300~Hz, 600~Hz, 900~Hz, and 1200~Hz). This signal simulates a typical variable-speed motor control sequence. Simultaneously, the EM injection carrier $V_{inj}(t)$ couples onto the high-current loop formed by the Drain-Source path and the load. 
The recovered spectrogram shown in~\cref{fig:feasibility test.} displays a clear ``staircase'' pattern corresponding to the frequency steps. Nevertheless, strong harmonics accompany each fundamental frequency.

\textbf{(4) Power Converter.}
The AC-DC rectification stage of power converters is the entry point of electrical energy for electronic devices, converting high-voltage AC mains into DC power. 
To investigate leakage from power supply units, we connected a power converter with a fixed high-power resistive load. Here, the target secret signal could be the AC mains voltage itself ($V_{sec}(t)$ at 50~Hz), representing the power consumption of the device. The measurement results (\cref{fig:feasibility test.}) reveal a prominent modulation effect: the fundamental mains frequency and a rich set of harmonics (e.g., 100~Hz, 150~Hz) are clearly recovered as sidebands around the carrier. This strong modulation arises from the bridge rectifier diodes at the adapter's input. Consistently, no signal appeared in the passively eavesdropped traces. 

\subsubsection{Leakage Characteristics}

After confirming the existence of injection-induced leakage, we further seek to characterize the quantitative relationships between the injection, leakage, and secret signals. We observe that injection-induced side channels introduce both the benefit of secret signal amplification and the challenge of nonlinear distortions. 

\begin{figure}[t]
    \centering
    \includegraphics[width=1\linewidth]{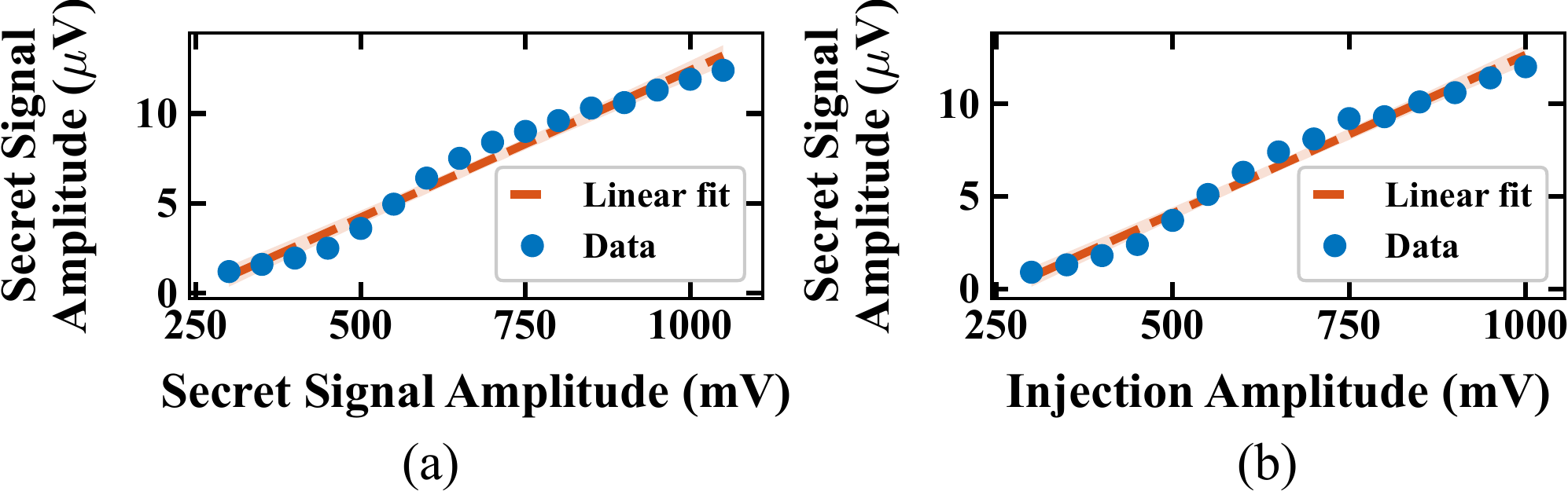}
    \caption{{\color{revcol}The amplitude of the leakage signal is jointly determined by the strength of the secret and injection signals.}}
    \label{fig:lamp_power_sideband}
\end{figure}

\textbf{Near-Linear Signal Amplification.}
We reused the setup above to produce different strengths of secret and injection signals to examine the quantitative relationship revealed by~\cref{eq:effi}. 
Our experiments confirm the near-linear relationship between $|V_{eav}(t)|$, $|V_{sec}(t)|$, and $|V_{inj}(t)|$. For example, \cref{fig:lamp_power_sideband} shows the variations of these quantities on the AD623 amplifier, where $|V_{eav}(t)|$ scales almost proportionally with $|V_{sec}(t)|$ and $|V_{inj}(t)|$. Results of the other three nonlinear components exhibit highly similar trends and are thus omitted. This result thus demonstrates the capability of injection-induced EM side channels in leaking fine-grained signals, where the eavesdropped signals can vary continuously according to the original analog secrets and the intended amplification controlled by the EM injection signal. 

\textbf{Challenge of Nonlinear Distortions.}
Although the linear amplification effect of the second-term inter-modulation provides injection-induced side channels with the unique capability of controllable leakage strength, it also inevitably faces the distortions caused by the higher-order terms ($V^{c}_{hi}(t)$ in~\cref{eq:Vmod}). This is illustrated by the nonlinear variations of the data points in~\cref{fig:lamp_power_sideband}, and could manifest as harmonics of the original secret signal  (e.g., as shown in~\cref{fig:feasibility test.}). While the harmonics are discernible when the secret is a simple single tone, they may degrade the quality of more complex wideband signals by contaminating the original frequency components. For example, human speech signals naturally consist of a fundamental component and its harmonics, meaning that higher-order inter-modulation products of the fundamental and even lower-order harmonics can overlap with, and thereby distort, higher-frequency components, posing a unique challenge for reconstructing high-quality audio and other wideband secrets.

%% file: sections/4-eavesdropping-design.tex
\subsection{InjectEave Design}
\label{sec:design}
Based on the new knowledge about the capability and challenge of injection-induced side channels, we provide an exemplary eavesdropping design, named \alias. 

\textbf{EM Injection and Receiving Hardware.}
Aiming for a portable and efficient design, \alias{} uses an Ettus USRP B210 and a Log-Periodic antenna to send single-tone injection signals at its maximum output power. 
On the leakage receiver side, another Log-Periodic antenna connected to a Siglent SSA3075X Plus spectrum analyzer collects the leaked electromagnetic signals. The spectrum analyzer demodulates the signal at the same center frequency as the USRP's output. The resulting baseband signal is routed to a recording device, such as a laptop, for further processing. The hardware setup is shown in~\cref{fig:experiment setup}.

{\color{revcol}
\textbf{Frequency Profiling.}
To identify effective injection carriers, we use a two-stage coarse-to-fine frequency profiling procedure. During this process, target devices are set to their typical operating states: audio devices continuously play a 2~kHz single-tone signal, smart fans operate at maximum speed, and smart lamps are profiled using the 50~Hz power-frequency component. We first conduct a coarse-grained frequency sweep from 70~MHz to 2~GHz with a 10~MHz step size to locate effective frequency ranges. Subsequently, a fine-grained frequency sweeping is performed within these effective ranges using a 1~MHz step. By comparing the leakage strength across all frequency candidates, we select the one with the highest leakage strength as the optimal carrier.
}

\textbf{Signal Enhancement Software.}
To address the observed distortions, especially for audio signal eavesdropping, we employ a Score-based Generative Model for Speech Enhancement (SGMSE)~\cite{welker22speech} backend and treat the noisy measurement as a structural anchor. 

The reconstruction is formulated as a mean-reverting diffusion process driven by two distinct mechanisms. The drift term uses the leakage as a structural constraint, locking the generation to the envelope $|V_{eav}(t)|$ to preserve the victim's original prosody. Simultaneously, the score function acts as a spectral enhancer, specifically filtering out the inter-modulation distortions $V^{c}_{hi}(t)$ to restore high-fidelity speech.

The effectiveness of the model relies on training with a large-scale synthesized dataset of paired clean and distorted signals. \rev{Given the practical difficulty of collecting aligned EM data in the wild, the speech enhancement model is trained solely on synthesized data}, with training pairs generated using a physics-based pipeline derived from~\cref{eq:Vmod}. We use LibriSpeech~\cite{panayotov2015librispeech} as the clean speech dataset and synthesize the corresponding leakage traces by \rev{applying various device-agnostic nonlinear coefficients to simulate the hardware nonlinearity of possible devices}, combined with ambient EM noise. \rev{Since physical devices differ in their nonlinear coefficients, the trained model is evaluated on unseen devices to assess its cross-device generalizability.} This approach ensures the model learns the unique spectral structure of these inter-modulation distortions $V^{c}_{hi}(t)$, enabling robust generalization to real-world hardware leakage.
{\color{revcol}
As demonstrated in~\cref{sec:case_studies}, enhanced audio exhibits significant improvements in both standard audio quality metrics and intelligibility. The audio demos can be found in~\cite{injecteave2025demo}.
}

%% file: sections/5-evaluation-in-a-laboratory-setting.tex
\section{Evaluation in a Laboratory Setting}
\label{sec:evaluations_in_lab}
This section evaluates \alias{} attack on 11 commercial devices of 5 different categories and measures the eavesdropping attack's performance in real-world settings.

\begin{figure}[t]
    \centering
    \includegraphics[width=0.75\linewidth]{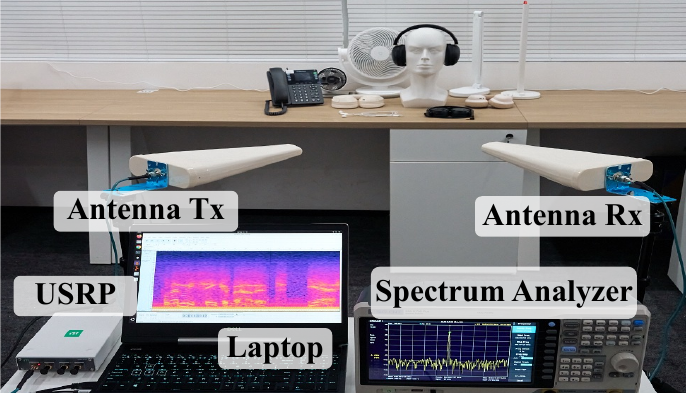}
    \caption{The hardware setup for evaluating COTS devices.}
    \label{fig:experiment setup}
\end{figure}

\subsection{Experimental Setup}
\label{sec:experiment_setup}

\input{tables/table1}

\cref{fig:experiment setup} shows the laboratory setup. The adversary's equipment consists of a USRP, directional Tx/Rx antennas, a spectrum analyzer, and a laptop. The laptop controls the USRP to generate the injected carrier, which is transmitted through the Tx antenna toward the target device. The Rx antenna captures the injection-induced EM leakage from the target, and the received signal is observed and measured by the spectrum analyzer. The evaluated COTS devices are placed on the desk as victim devices during the experiment.

\textbf{Victim Devices.} 
The evaluated devices included: (1) three wired headphones from Sony, Dell, and Apple; (2) three wireless headphones from UGreen, PHILIPS, and HP; (3) a wireless landline from Flyingvoice; (4) two smart fans from Xiaomi and OIDIRE; and (5) two smart lamps from Xiaomi and JINGZAO. The detailed information of each device is specified in~\cref{tab: Attacks on COTS Devices}.

\textbf{Metrics.} We evaluate the overall performance of \alias{} attack on the signal, feature, and semantic levels respectively, using two primary metrics:

(1) \textit{Signal-to-Noise Ratio (SNR)} characterizes the signal-level quality of the injection-induced EM leakage and the recovered signal at specific distances.
(2) \textit{Attack Success Rate (ASR)} provides a unified measure of recovered information fidelity relative to the ground truth. For audio devices, ASR is defined as $1 - WER$, where word error rate (WER) evaluates the accuracy of recovering a speech signal's semantic information. {\color{revcol} We compute WER by first transcribing the recovered speech using Whisper~\cite{radford2023whisper}, and comparing the transcript with the ground-truth script at the word level, where substituted, missing, and extra words are counted as errors.} For discrete-state devices, including smart fans and smart lamps, ASR is defined as the classification accuracy ($N_{\text{correct}} / N_{\text{total}}$) across all operational states.

\subsection{Evaluations on COTS Devices}
We categorize the analysis on the 11 COTS devices into two distinct threat dimensions based on the attack surfaces and exposed privacy risks: high-fidelity audio recovery and human activity inference. Specifically, audio peripherals serve as direct attack vectors for speech eavesdropping. In contrast, smart home devices' states can be exploited to infer user presence and behavioral patterns. 

{\color{revcol} 
To systematically evaluate the attack performance, we first conduct measurements at 50~cm across all devices, quantifying the injection frequency, SNR, and signal recognition rate, before further evaluating the maximum achievable attack distance. 
\cref{tab: Attacks on COTS Devices} reports the SNR and recognition rate at 50~cm. We establish a device-specific SNR threshold from 30 trials, set to the lowest observed SNR, and then conduct another 30 independent trials, counting a trial as successful if the measured leakage SNR exceeds this threshold. For maximum distance evaluation, the detection SNR threshold is defined as $10\log_{10}(10/B_n)$, derived from a noise-only spectrum with a 10~Hz resolution bandwidth, where $B_n$ is the observation bandwidth. For audio devices (wired headphones, wireless headphones, and landlines), we set $B_n=20$~kHz to cover the audible speech band; for fans and lamps, $B_n=150$~Hz, covering three times the leakage frequency. These bandwidths yield detection thresholds of $-33.0$~dB for audio devices and $-11.8$~dB for fans and lamps.}

\subsubsection{Threats Against Audio Peripherals}
Audio peripherals, including wireless headphones, wired headphones, and landlines, are ubiquitous in both public and private spaces. To evaluate \alias{} attack on these COTS audio peripherals, we play a 2~kHz reference single-tone signal, which represents a typical speech signal frequency, on the devices.  As shown in~\cref{tab: Attacks on COTS Devices}, \alias{} achieves a near 100\% signal recognition rate across these devices, with each device exhibiting injection-sensitive frequency ranges that allow flexible carrier selection.

\textbf{Wired Headphone.}
We evaluate \alias{} attack on wired headphones to demonstrate effectiveness across different host-device interfaces. As summarized in~\cref{tab: Attacks on COTS Devices}, we achieve a near 100\% recognition rate at 50~cm across all setups. 
The optimal injection frequency for the Sony ZX110AP exhibits a significant host-dependent shift, requiring a distinct frequency band when connected with the MacBook Pro M2 compared to the Dell G5.

Despite these variations, the attack remains robust with effective ranges of 5~m and 4~m, respectively. In contrast, the Apple Earbuds on an iPhone 15 Pro achieves an SNR of 5.9~dB and a 1~m range. These results indicate that the analog amplifier of the host device's internal sound card is the fundamental component exploited for injection-induced EM leakage, while the headphone cable primarily acts as an unintentional antenna. The distinct grounding and circuit layouts of the Dell and MacBook sound cards determine the specific resonant frequencies required for inter-modulation, resulting in the phenomenon that the same headphone is susceptible at different injection frequency bands.

\textbf{Wireless Headphone.} 
We evaluate three wireless models, each exhibiting high susceptibility to \alias{} across distinct frequency bands as detailed in~\cref{tab: Attacks on COTS Devices}. The UGreen MAX2 demonstrates the most robust leakage, achieving an SNR of 23.1~dB and a 100\% recognition rate with an effective distance of up to 6~m. Similarly, the PHILIPS TAH2020 supports a 100\% recognition rate with an SNR of 20.9~dB and a 6~m range. Even the HP H231R maintains a 29/30 recognition rate and a 4~m distance.
We further validate the real-world threat of \alias{} in~\cref{sec:casestudy1}, where we achieve successful through-wall audio eavesdropping in real-world hotel and conference room scenarios.

\textbf{Landline.} Landline desk phones are often used in high-sensitivity scenarios such as government offices for internal communications~\cite{kuhn2005security}. To examine risks in these professional environments, we extend our evaluation to a Flyingvoice P23GW VoIP landline. We simulate an active call and sweep the carrier frequency to locate sensitive nonlinear junctions within the handset's internal amplifier and ADC stages. As summarized in~\cref{tab: Attacks on COTS Devices}, the system exhibits high susceptibility, achieving a 100\% recognition rate and an SNR of {\color{revcol} 12.9~dB}. The attack remains viable at distances up to 3~m, allowing an adversary to eavesdrop on confidential corporate negotiations without compromising the digital network. We further demonstrate the severity of this threat in~\cref{sec:casestudy3}, where \alias{} is used to achieve closed-loop manipulation of the landline's voice interface.

\textbf{Effect of Audio Signal Enhancement.}
We further examine the effectiveness of the eavesdropping design in~\cref{sec:design} on 2 minutes of speech audio recordings captured at a distance of 50~cm with 65~dB volume from the UGreen MAX2 wireless headphone using both SNR and Short-Time Objective Intelligibility (STOI), a standard speech quality metric. The results show notable improvements in both STOI and SNR. The average SNR increases drastically from 7.0~dB to 16.1~dB, indicating that the model successfully suppressed the dominant background noise and the injection-induced carrier phase noise. Furthermore, STOI sees a significant boost from 0.58 to 0.72. Given that STOI is highly correlated with human speech intelligibility, this improvement confirms that our algorithm effectively reconstructs the phonetic details masked by the hardware's nonlinear harmonics, rendering the eavesdropped speech intelligible to human listeners. 
Our demos can be found in~\cite{injecteave2025demo}.

{\color{revcol}
\textbf{Cross-device Profiling Transferability.} We evaluate cross-device profiling transferability using three UGreen MAX2 headphones of the same model. For each headphone, we profile its effective carrier frequency and directly reuse it to attack the other two without re-profiling. 
All cross-device profiling attacks succeed in our experiments. The transferred attacks achieve leakage SNRs of 23.2--24.6~dB, with only 0.8--2.9\% relative deviation compared with attacks using device-specific profiling. This good profile transferability mainly stems from the fact that the profiled frequency (942~MHz) is identical across three devices and also indicates that the effectiveness of transferability is largely determined by the shared analog front-end layout, cabling structure of the same model, rather than by batch or manufacturing differences.
}

\subsubsection{Threats Against Smart Home Appliances}
Beyond audio-centric devices, we examine the generalizability of the attack on IoT smart appliances, which are deeply integrated into private environments. The main leakage source shifts from audio amplifiers to the switching MOSFETs in fans and power converters in lamps.

\begin{figure}[t]
    \centering
    \includegraphics[width= 1.0\linewidth]{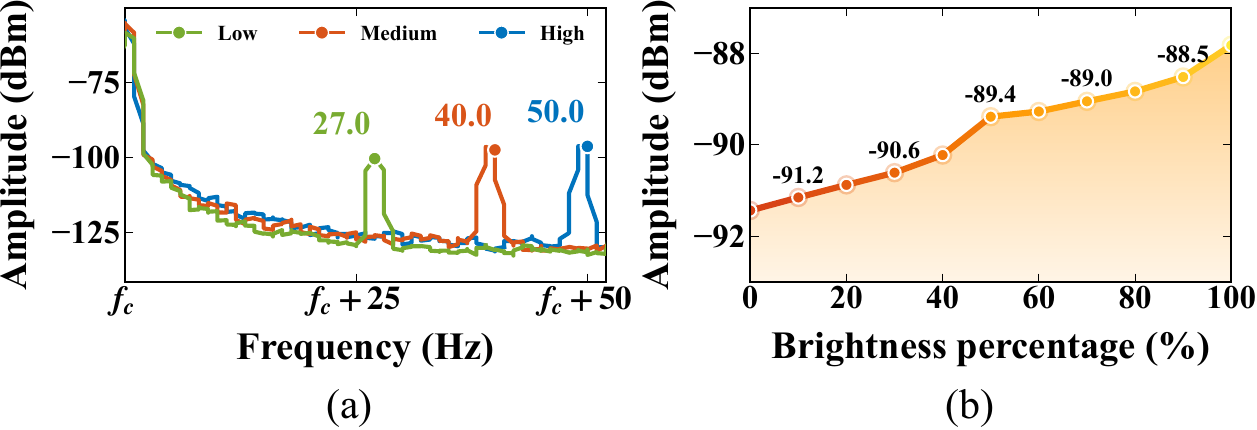}
    \caption{Eavesdropping on smart home activities: (a) inferring fan speed, and (b) inferring lamp brightness.}
    \label{fig:fan_and_lamp}
\end{figure}

\textbf{Smart Fan.} The OIDIRE ODI-MF10A and Xiaomi BPLDS10DM smart fans feature operational modes—such as Sleep, Natural, and Standard—that serve as proxies for the user's activity. The attack exploits the modulation caused by the motor's driving signal, where the rotational speed (e.g., low: 27~Hz, medium: 40~Hz, high: 50~Hz) modulates the injected carrier. As shown in~\cref{tab: Attacks on COTS Devices}, the OIDIRE model yielded an SNR of \rev{38.6~dB} at 480~MHz, maintaining a successful recognition rate up to 6~m. Unlike audio devices, the fan's leakage manifests as sidebands at distinct frequency offsets in the frequency domain (\cref{fig:fan_and_lamp}~(a)). This enables ``Context Inference'' attacks: for example, detecting ``Sleep Mode'' at night can confirm a user's rest schedule and infer occupancy without the need for visual surveillance.

\textbf{Smart Lamp.} We further tested the Xiaomi 1S and JD JINGZAO JDO-06 smart lamps. The mechanism exploits the nonlinearity of the lamp's power converter, where the 50~Hz AC mains current modulates the injected carrier. As indicated in~\cref{tab: Attacks on COTS Devices}, the Xiaomi 1S exhibited strong leakage with an SNR of 21.6~dB. Since the amplitude of the demodulated signal is proportional to the power load, we can remotely infer the precise dimming level. By mapping these power levels to vendor-specific presets (e.g., 20\% brightness for ``Reading''), an adversary can perform ``Behavioral Profiling,'' transforming a simple light source into a beacon that exposes a user's specific activities and routines without requiring any network-level access. \cref{fig:fan_and_lamp}~(b) demonstrates how adversaries could precisely infer the lamp's brightness from the strength of received signals, effectively enabling power side-channel analysis~\cite{narimani2024exploring, bursztein2023generalized} in a contactless manner. 

\subsection{Environmental Impact Quantification}
\label{sec: Impact Quantifications}

To characterize the physical limits and practical constraints of \alias{} attack, we select three representative devices for impact quantification: the UGreen MAX2 wireless headphones, the OIDIRE ODI-MF10A smart fan, and the Xiaomi 1S smart lamp. Unless otherwise specified, our default experimental configuration employs a 50~cm attack distance, a 90$^\circ$ antenna orientation, and an injection power of 18~dBm. The carrier frequencies are configured at 940~MHz, 480~MHz, and 100~MHz for wireless headphones, smart fan, and smart lamp, respectively, corresponding to their optimal resonant points discovered during the frequency sweep experiment. We quantify the attack robustness by varying the attack distance to determine the effective range, adjusting the antenna orientation to analyze polarization sensitivity, and introducing various material barriers to evaluate signal penetration and eavesdropping in realistic environments.

\begin{figure}[t]
    \centering
    \includegraphics[width=0.9\linewidth]{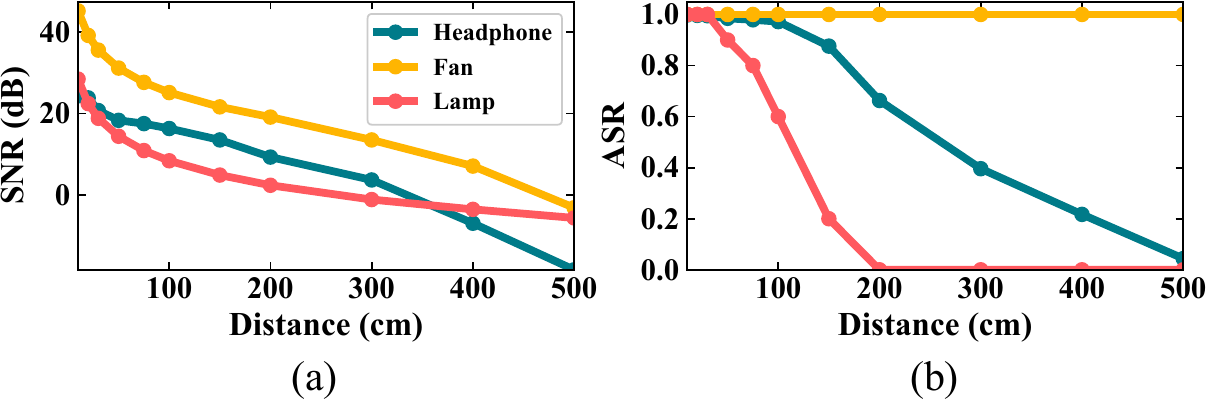}
    \caption{Impact of attack distance on (a) SNR trend and (b) Attack Success Rate (ASR) across different COTS devices.}
    \label{fig:impact_distance}
\end{figure}

\textbf{Impact of Antenna-Target Distance.}
To evaluate the effective eavesdropping distance and the limits imposed by propagation path loss, we varied the distance between the adversary and the victim device up to 5~m. As shown in~\cref{fig:impact_distance}~(a), the leakage SNR decreases monotonically with distance, dropping by approximately 45~dB across all devices from 10~cm to 5~m.
The corresponding ASR results in~\cref{fig:impact_distance}~(b) reveal distinct hardware-level resilience: while the Xiaomi 1S smart lamp becomes resilient beyond 2~m and the UGreen MAX2 drops to a 4.4\% ASR at 5~m, the OIDIRE smart fan maintains a 100\% ASR.
The attack distance can be further extended by increasing transmitting power or lowering the phase noise by using high-performance attack equipment. Replacing the Siglent SSA3075X Plus (phase noise: -98~dBc/Hz) with a high-end Keysight N9000B spectrum analyzer (phase noise: -110~dBc/Hz), which offers a superior phase noise of -110~dBc/Hz at a 1~GHz carrier with 10~kHz offset, extends the effective distance to over 8~m for UGreen wireless headphone, 10~m for OIDIRE smart fan, and 5~m for Xiaomi smart lamp. These findings highlight that \alias{} poses a significant long-range threat in practical environments by eliminating the requirement for physical proximity.
We provide three case studies in~\cref{sec:case_studies} to show the real-world threat of \alias{} attack in the wild.

\begin{figure}[t]
    \centering
    \includegraphics[width=.8\linewidth]{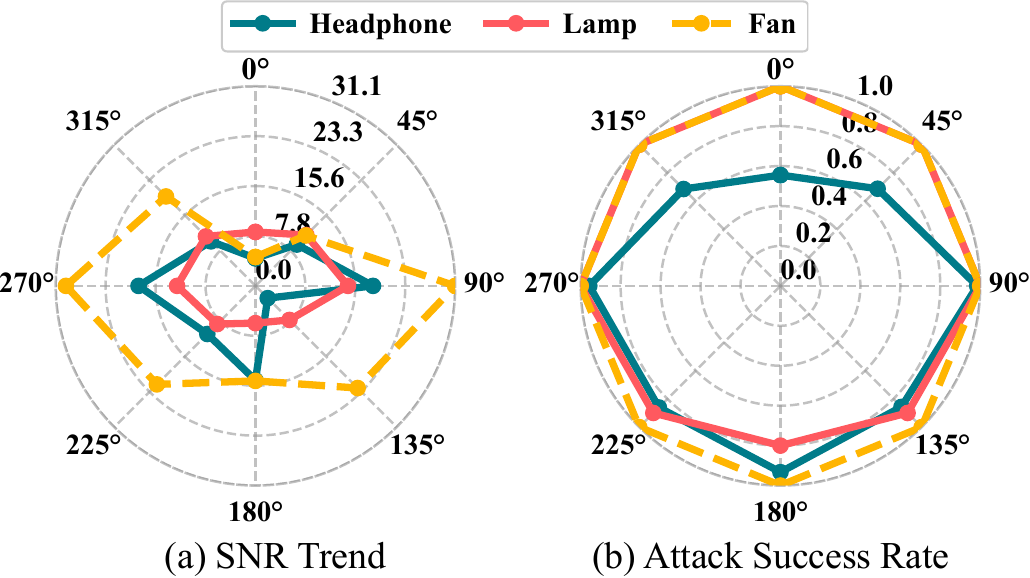}
    \caption{\alias{}'s robustness against antenna angles. The results indicate optimal performance at 90$^\circ$.}
    \label{fig:impact_angle}
\end{figure}

\textbf{Impact of Antenna Angle.}
To investigate the impact of the relative angle between the transmitting and receiving antennas, we rotated the receiving antenna along the target's azimuthal plane from $0^\circ$ to $315^\circ$ in $45^\circ$ increments while keeping the transmitting antennas fixed. As shown in~\cref{fig:impact_angle}~(a), the leakage exhibits strong directionality, with all devices peaking at $90^\circ$.
At this optimal orientation, the SNR reaches 18.3~dB, 31.1~dB, and 14.4~dB for the headphone, fan, and lamp, respectively. This phenomenon occurs because a $90^\circ$ spatial separation maximizes isolation between the antennas, effectively suppressing direct carrier leakage into the receiver. This prevents receiver saturation and lowers the noise floor, thereby maximizing the SNR of the recovered side-channel signal.
The ASR in~\cref{fig:impact_angle}~(b) largely follows the SNR trends but reveals a task-specific resilience. For the UGreen headset, the audio eavesdropping ASR fluctuates, peaking at 93.7\% ($90^\circ$) from a low of 50.5\% ($0^\circ$). 
We observe that attack performance is maximized when aligning the receiving antenna near the $90^\circ$ orientation, providing insightful guidance for practical attack design and deployment. Nevertheless, precise alignment to $90^\circ$ is not strictly required in practice, as the recognition rate remains above 86\% even with a $15^\circ$ deviation from the optimal $90^\circ$ orientation. 

In contrast, state-based inference is remarkably robust: the OIDIRE smart fan maintains a consistent 100\% ASR in all orientations, while the Xiaomi smart lamp sustains a high ASR with only minor fluctuations between $135^\circ$ and $225^\circ$.

\begin{figure*}[!t]
    \centering
    \includegraphics[width=1\linewidth]{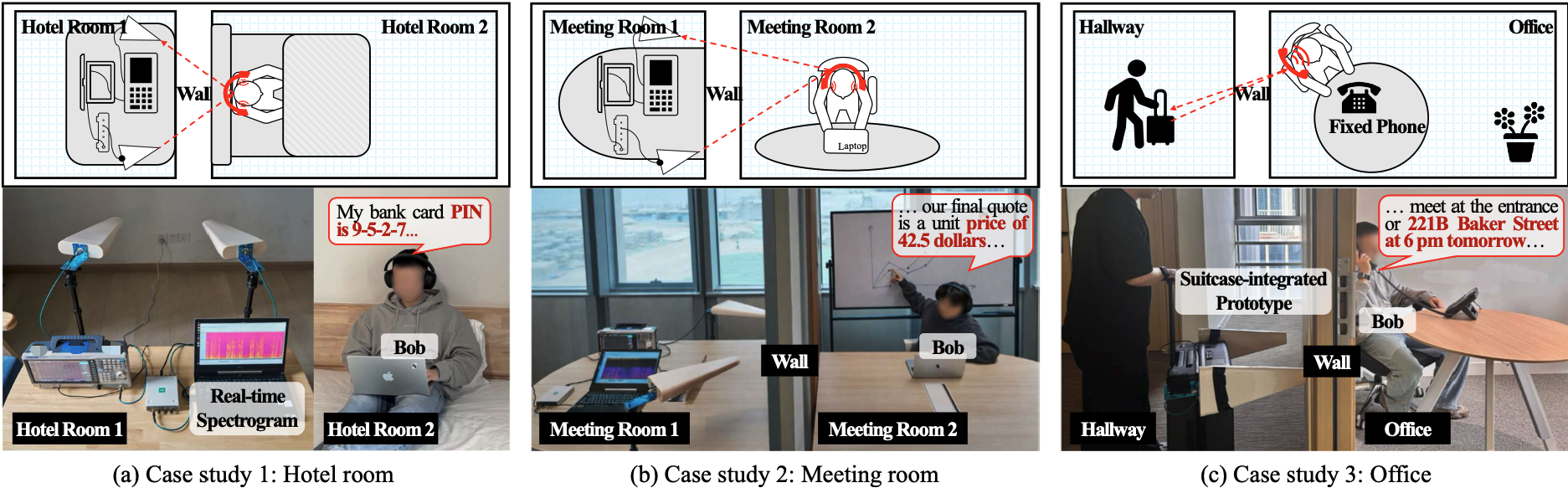}
    \caption{\protect\centering Case studies of representative audio eavesdropping scenarios in the wild.}
    \label{fig:casestudies}
\end{figure*}

\begin{figure}[t!]
    \centering
    \includegraphics[width=\linewidth]{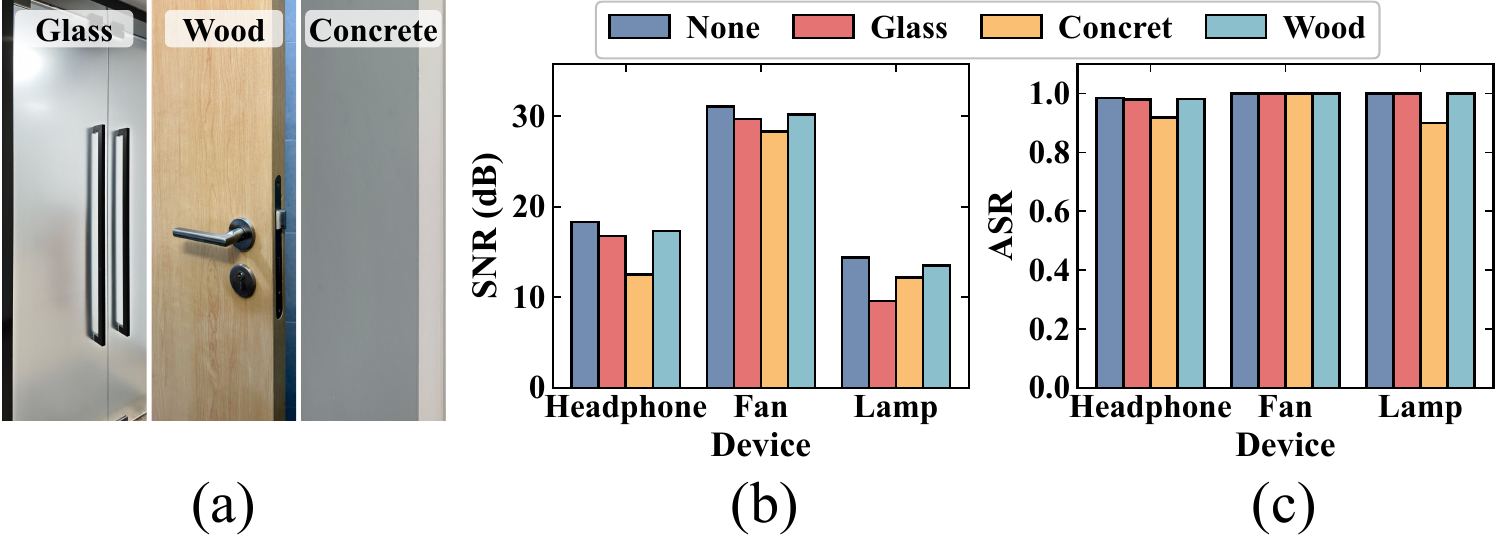}
    \caption{Impact of different barriers on \alias{} performance, showing feasibility of through-wall eavesdropping in real-world non-line-of-sight (NLoS) environments.}
    \label{fig:impact_barrier}
\end{figure}

\textbf{Impact of Physical Barrier.}
To evaluate \alias{} in non-line-of-sight (NLoS) environments, we measured the signal attenuation caused by common structural materials, including glass, wood, and concrete. As shown in~\cref{fig:impact_barrier}~(b), these barriers exert minimal influence on the leakage SNR. Glass and wood induce a negligible attenuation of only 1–2~dB compared to line-of-sight (LoS) conditions. Concrete obstacles cause a more pronounced but still limited drop: 5.8~dB for the headphone, and approximately 2.5~dB for the fan and lamp.
The ASR remains remarkably resilient across all tested materials in~\cref{fig:impact_barrier}~(c). For glass and wood, the ASR remains unchanged for all devices. Even with concrete, the impact is marginal: the headphone ASR decreases by only 3\%, while the smart fan maintains a 100\% success rate. These results demonstrate that \alias{} effectively penetrates common structural barriers, enabling covert through-wall eavesdropping in partitioned indoor environments such as offices and hotels, as further detailed in~\cref{sec:case_studies}.

%% file: tables/table1.tex
\begin{table*}[t]
\centering
\footnotesize
\renewcommand{\arraystretch}{0.85}

\begin{threeparttable}

\caption{Summary of Attacks on COTS Devices}
\label{tab: Attacks on COTS Devices}

\begin{tabular*}{\textwidth}{@{\extracolsep{\fill}}lllcccccc}
\toprule

\textbf{Device Type} & \textbf{Brand} & \textbf{Model} & \textbf{Year} & \textbf{\makecell[c]{Source\\of leakage}} & \textbf{\makecell[c]{Injection\\Frequency }} & \textbf{\makecell[c]{SNR \tnote{\ddag} }} & \textbf{\makecell[c]{Recog.\\Rate \tnote{\S}}} & \textbf{\makecell[c]{Max\\Dist. \tnote{\P}}} \\
\midrule

\multirow{3}{*}{\textbf{\makecell[l]{Wired\\Headphones}}}
& Sony + Dell\tnote{$\|$} & ZX110AP & 2014 & \multirow{3}{*}{Amplifier} & *1*--*7*\,MHz & {\color{revcol}$21.7 \pm 0.8$\,dB} & 30/30 & 5\,m \\ \cmidrule{2-4} \cmidrule{6-9}
& Sony + Mac\tnote{$\|$} & ZX110AP & 2014 & & *2*--*6*\,MHz & {\color{revcol}$13.9 \pm 0.6$\,dB} & 30/30 & 4\,m \\ \cmidrule{2-4} \cmidrule{6-9}
& Apple + iPhone\tnote{$\|$}& Earbuds & 2016 & & *6*--*8*\,MHz & {\color{revcol}$5.9 \pm 0.5$\,dB} & 29/30 & 1\,m \\
\midrule

\multirow{3}{*}{\textbf{\makecell[l]{Wireless\\Headphones}}}
& UGreen\tnote{\dag} & MAX2 & 2024 & \multirow{3}{*}{Amplifier} & *0*--*8*\,MHz & {\color{revcol}$23.1 \pm 0.7$\,dB} & 30/30 & 6\,m \\ \cmidrule{2-4} \cmidrule{6-9}
& PHILIPS & TAH2020 & 2025 & & *9*--*5*\,MHz & {\color{revcol}$20.9 \pm 0.7$\,dB} & 30/30 & 6\,m \\ \cmidrule{2-4} \cmidrule{6-9}
& HP & H231R & 2023 & & *8*--*2*\,MHz & {\color{revcol}$19.9 \pm 0.8$\,dB} & 29/30 & 4\,m \\
\midrule

\textbf{\makecell[l]{Landline}} & Flyingvoice & P23GW & 2023 & \makecell[c]{ADC \&\\Amplifier} & *4*--*2*\,MHz & {\color{revcol}$12.9 \pm 0.6$\,dB} & 30/30 & 3\,m \\
\midrule

\multirow{2}{*}{\textbf{Fan}}
& OIDIRE\tnote{\dag} & ODI-MF10A & 2023 & \multirow{2}{*}{\makecell{Switching\\MOSFET}} & *4*--*2*\,MHz & {\color{revcol}$38.6 \pm 0.5$\,dB} & 30/30 & 6\,m \\ \cmidrule{2-4} \cmidrule{6-9}
& Xiaomi & BPLDS10DM & 2025 & & *3*--*1*\,MHz & {\color{revcol}$30.0 \pm 0.5$\,dB} & 30/30 & 4\,m \\
\midrule

\multirow{2}{*}{\textbf{Lamp}}
& JINGZAO & JDO-06 & 2024 & \multirow{2}{*}{\makecell{Power\\Converter}} & *8*--*5*\,MHz & {\color{revcol}$20.0 \pm 0.7$\,dB} & 29/30 & 3\,m \\ \cmidrule{2-4} \cmidrule{6-9}
& Xiaomi\tnote{\dag} & 1S & 2019 & & *7*--*0*\,MHz & {\color{revcol}$21.6 \pm 0.8$\,dB} & 30/30 & 3\,m \\
\bottomrule

\end{tabular*}

\begin{tablenotes}[para]
\footnotesize
\scriptsize
$\|$: Three different host devices for wired headphones. \ $\ast$: We intentionally hide the injection frequency for ethical considerations. \ \dag: Devices evaluated in~\cref{sec: Impact Quantifications}. \ \ddag: {\color{revcol}SNR is evaluated at 50\,cm distance and reported as mean $\pm$ standard deviation across 30 independent trials.} \ \S Recognition rate is evaluated at 50\,cm distance. \ \P Max distance is evaluated at 5/30 recognition rate.
\end{tablenotes}

\end{threeparttable}
\end{table*}

%% file: sections/6-audio-eavesdropping-in-the-wild.tex
\section{Audio Eavesdropping in the Wild}
\label{sec:case_studies}

This section moves beyond controlled laboratory characterization to demonstrate end-to-end audio eavesdropping in realistic environments. Our case studies in~\cref{fig:casestudies} focus on non-line-of-sight (NLoS) scenarios where the adversary is physically separated from the victim and performs through-wall attacks against victim-side audio devices. By launching \alias{}, the adversary can recover a remote participant's voice without physical access to the speaker. We further examine practical deployment factors, including playback volume, nearby active electronics, ambient RF interference, and injection stealthiness. Finally, we demonstrate two extended threats: a closed-loop attack that synthesizes and injects recovered speech back into the victim device, and longer-distance eavesdropping enabled by an external power amplifier.

\subsection{Real-World Audio Eavesdropping}

\label{sec:casestudy1}
This case study demonstrates that the injection-induced side channel can be exploited to conduct through-wall eavesdropping to compromise individual privacy (e.g., bank PINs, personal agendas) and corporate confidentiality (e.g., procurement quotes, strategic timelines). 
As shown in~\cref{fig:casestudies} (a) and (b), we conduct end-to-end attacks in both a hotel room and a meeting room setting. In these scenarios, the victim Bob engages in a confidential video call using a UGreen MAX2 wireless headset. The adversary operates from an adjacent room, separated by a 30~cm solid concrete wall, with a straight-line distance of over 1~m.
We use a commercial open-source text-to-speech (TTS) tool~\cite{ttsmaker2024} to synthesize six personal conversation segments in both male and female voices, covering personal and business-oriented dialogues. The content of the conversation segments is provided in~\cite{injecteave2025demo}.

\begin{figure}[t]
    \centering
    \includegraphics[width=0.9\linewidth]{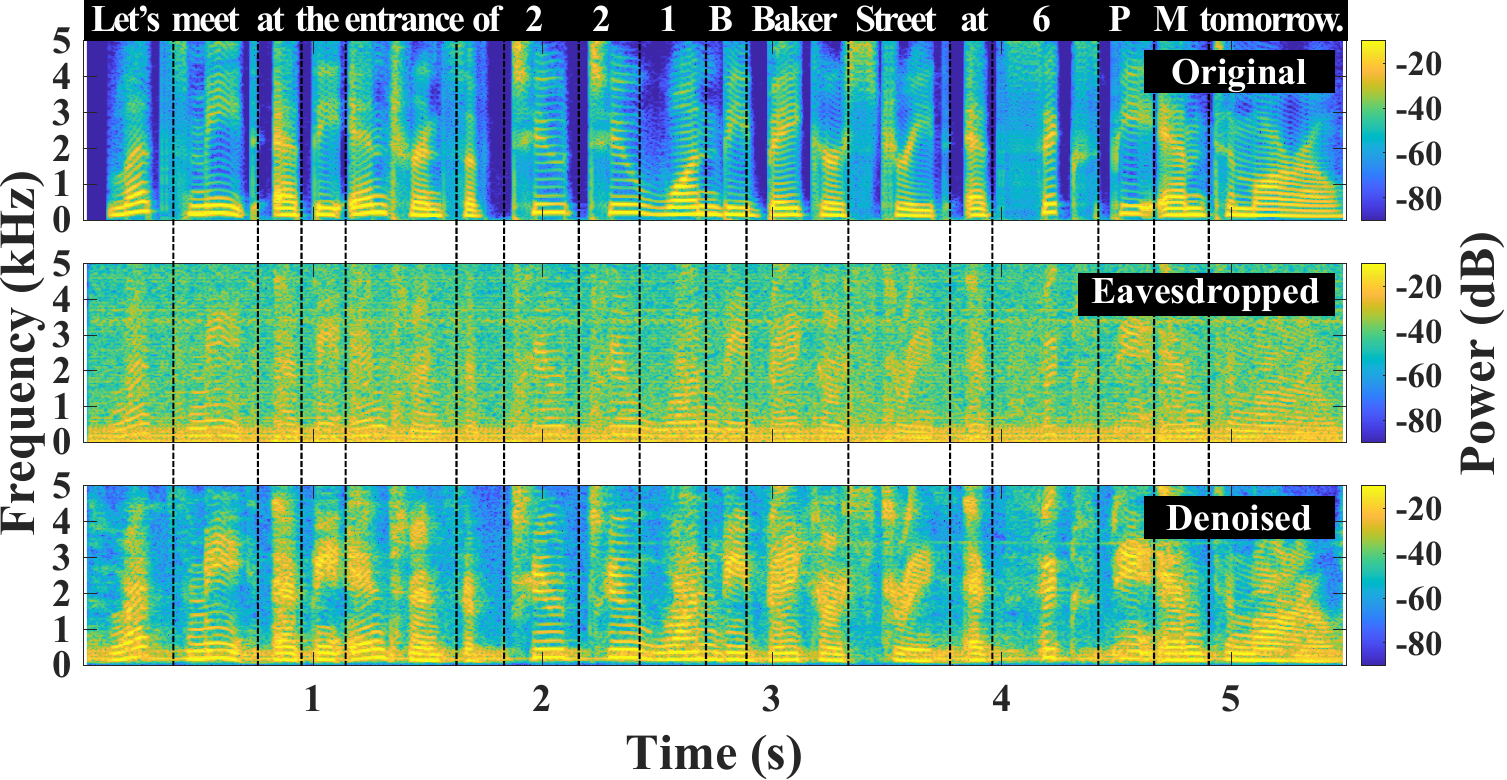}
    \caption{Comparison of audio spectrograms across the original, eavesdropped and denoised audio, demonstrating the eavesdropping performance and signal enhancement effect in \alias{} attack.}
    \label{fig:stft_comparition}
\end{figure}

\cref{fig:stft_comparition} shows a representative time-frequency analysis of the eavesdropped audio, successfully recovering a speech segment of the victim's detailed personal agenda: ``Let's meet at the entrance of 221B Baker Street at 6 PM tomorrow.'' The top panel displays the clean ground-truth audio for reference. Despite the significant attenuation caused by the solid concrete wall, the eavesdropped spectrogram (middle panel) clearly retains the fundamental harmonic structures of the original speech. Applying the signal enhancement algorithm described in~\cref{sec:design} significantly sharpens spectral harmonics while suppressing background interference (bottom panel). These results demonstrate that the injection-induced EM side channel enables reliable eavesdropping of Alice's intelligible speech even through dense physical barriers, posing direct threats to private conversations.

{\color{revcol} 
To further evaluate the robustness of \alias{} attacks across diverse real-world teleconferencing conditions, we examine three factors that may affect attack performance: the headphone playback volume, interference from nearby active electronics and ambient RF noise, and the perceptibility of the injected carrier at the victim device. 
}

\begin{figure}[t]
    \centering
    \includegraphics[width=0.7\linewidth]{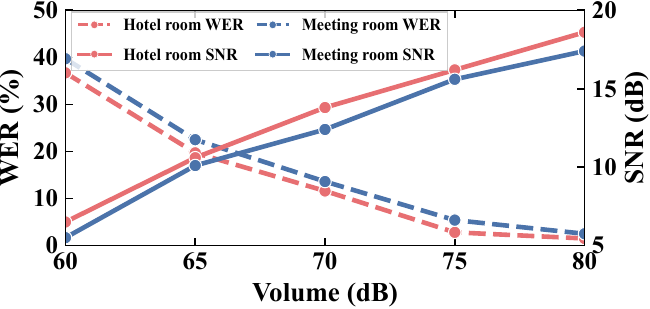}
    \caption{Impact of audio volume on SNR and word error rate of eavesdropped speech signals.}
    \label{fig:impact_volume}
\end{figure}

{\color{revcol} 
\textbf{Evaluation of Headphone's Playback Volume.}
}
We adjust the headphone's playback volume from a quiet office level of 60~dB to a louder entertainment level of 80~dB. As shown in~\cref{fig:impact_volume}, the SNR of the eavesdropped leakage exhibits a strong positive correlation with the playback volume. Crucially, even at a modest sound pressure level (SPL) of 65~dB, a typical threshold for private business conversations~\cite{pearsons1977speech}, the proposed \alias{} attack remains highly effective.
Specifically, the adversary can successfully recover Alice's managerial instructions with an SNR of approximately 10~dB and a corresponding word error rate (WER) of roughly 22\%. These metrics indicate that even at lower volumes, the attack retains sufficient phonetic information to reconstruct Alice's intelligible speech.

\begin{figure}[t]
    \centering
    \includegraphics[width=0.8\linewidth]{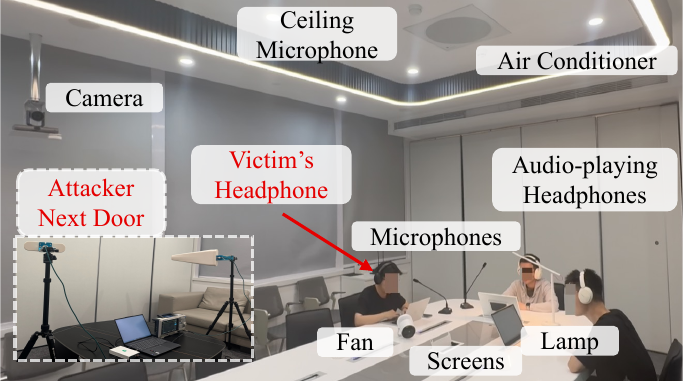}
    \caption{{\color{revcol}An office eavesdropping setting with diverse interference from multiple ceiling-mounted and desktop electronics. The injection-induced leakage remains robust to interference.}}
    \label{fig:app_multi_device_rf}
\end{figure}

{\color{revcol}
\textbf{Evaluation of Multi-device and Ambient RF Interference Effects.}
To evaluate whether \alias{} remains effective in a realistic environment with multiple nearby nonlinear devices and ambient RF activity, we conduct an experiment in a meeting room with diverse electronic devices, including ceiling cameras, ceiling microphone arrays, and a central air conditioner. The attack is launched on a commercial headphone while multiple nearby electronic devices operate simultaneously, including audio devices, microphones, and common household and office appliances. These devices are wirelessly connected via BLE and WiFi, representing a noisy RF environment. We measure the SNR at a 50~cm distance with the interference devices turned on versus off, and observe a deviation of only 2.2~dB, indicating that the attack remains stable in the presence of nearby active electronics and ambient RF activity. A photo and full device details are provided in~\cref{fig:app_multi_device_rf}. 

This robustness against interference from nearby devices is further explained by the frequency and spatial selectivity of \alias{}, as demonstrated on our website~\cite{injecteave2025demo}. Devices of different models often exhibit distinct effective frequencies, allowing the adversary to tune the injected carrier toward the target device while avoiding comparable responses from nearby non-target devices. When frequency profiles overlap for devices of the same model, antenna realignment provides spatial selectivity and keeps the target leakage dominant. 

\textbf{Evaluation of Injection Perceptibility.} The injected carrier may induce additional voltage variations in the target circuit. To assess whether this injected energy will cause humanly detectable artifacts in the output audio, we specifically place the transmitting antenna adjacent to the target UGreen MAX2 headphone and inject the single-tone carrier at 18~dBm power, representing a worst-case scenario with large injected energy, while ensuring this setting can induce recoverable leakage. During the test, the headphone plays normal audio, and we use an external microphone to directly record the audio output from the headphone under two conditions: with and without the injected carrier. The STOI deviation ranges from 0.001 to 0.012, with an average of 0.008. Demos~\cite{injecteave2025demo} also show no perceptible difference in the headphone output, suggesting that the single-tone carrier does not introduce noticeable audible artifacts or distort the victim's received audio even under aggressive near-field injection. The stealthiness is preserved because single-frequency carriers are filtered by analog front-ends, introducing little perceptible distortion.}

\subsection{Closed-Loop Conversation Manipulation}
\label{sec:casestudy3}

This case study demonstrates the real-world threat of \alias{} attack against critical office landline infrastructure, specifically focusing on the Flyingvoice P23GW VoIP landline phone in a private office environment. Beyond passive eavesdropping, \alias{} enables a novel ``Eavesdrop-Synthesize-Inject'' closed-loop manipulation chain that elevates the threat from compromising conversation confidentiality to falsifying speech content. As shown in~\cref{fig:casestudies}~(c), the victim, Bob, is at his desk and engaged in a call with his boss, Alice. The adversary packs the portable attack hardware into a suitcase-based prototype and stands in the hallway, maintaining a standoff distance of 50~cm from the target device, with a 20~cm office wall separating them. The internal structure of this prototype is illustrated in~\cref{fig:prototype}.
{\color{revcol}
The specific three steps for carrying out the closed-loop conversation manipulation are described as follows. 
}

\textbf{Step 1: Eavesdrop.} The attack starts by eavesdropping on the conversation through the injection-induced side channel to achieve context awareness. By eavesdropping on Alice's voice from the landline's speaker at 880~MHz, the adversary's system achieves a real-time understanding of the call. 

\textbf{Step 2: Synthesize.} Alice's recovered speech from the eavesdropping stage serves as the speaker reference for the voice-cloning module. Upon detecting predefined trigger keywords (e.g., ``quote'' or ``confirmation''), the system activates a voice-cloning module, IndexTTS-2~\cite{zhou2025indextts2} to generate a contextually appropriate and identity-specific deepfake response in real time. To preserve stealthiness, the adversary adjusts the synthesized speech using the eavesdropped audio as a reference, matching Alice's call audio in speech quality and perceived volume before modulation.

\textbf{Step 3: Inject.} The injected speech is amplitude-modulated onto a 1075~MHz carrier and transmitted at 40~dBm using well-established EM signal injection techniques~\cite{kune2013ghost, tu2018injected} through a separately profiled EM injection channel. The carrier then couples into the landline's analog audio output path and is demodulated by the nonlinear front-end into audible speech at Bob's headset. To avoid overlap between injected and received voice signals, Step 1 and Step 3 are designed to be time-divided: the adversary stops eavesdropping and switches to speech injection, injecting the synthesized Alice's audio into the target landline's audio output to complete the closed-loop manipulation. We conduct a series of trials across diverse semantic contexts to demonstrate the real-world attack impact of this closed-loop manipulation, \rev{with demos provided on our project website~\cite{injecteave2025demo}}.

{\color{revcol} To quantify the perceptibility of the injected audio manipulation, we record the victim-side audio using an external microphone. Specifically, under the same recording setup, we compare Alice's original speech directly played through Bob's headset with the corresponding audio injected into Bob's headset. 
The resulting STOI deviation is only 0.071 on average, indicating that the injected audio remains highly similar to the original speech of Alice as actually heard by Bob. Further demos can be found on \cite{injecteave2025demo}.}

\subsection{Long Distance Eavesdropping}
\label{sec: longer distance eavesdropping}
Attack distance directly affects the practical risk of InjectEave, as it determines whether an attacker can recover the victim's low-frequency secret without close physical proximity. Based on the leakage model in~\cref{eq:effi} and the theory of signal attenuation in free space~\cite{tse2005fundamentals}, we derive the theoretical maximum eavesdropping distance as: $d_{max} \propto \sqrt{|V_{inj}| \cdot |V_{sec}| / N_{sys}}$, where $V_{inj}$ denotes the voltage amplitude of the injected carrier, indicating that the range can be extended through higher transmission power, such as utilizing amplifiers.

\begin{figure}[t]
    \centering
    \includegraphics[width=0.95\linewidth]{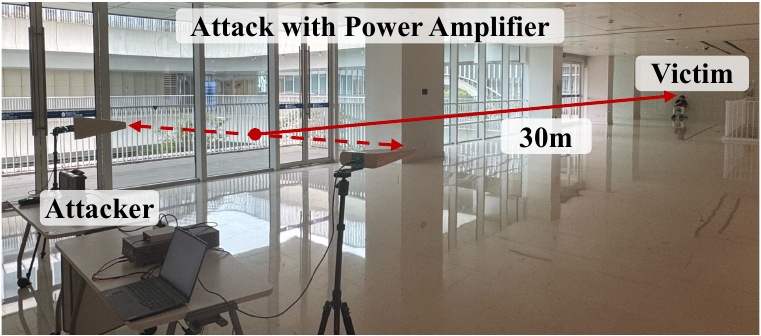}
    \caption{{\color{revcol}Long-distance eavesdropping setup with external power amplification, showing extended attack ranges of up to 30~m under higher EM injection power.}}
    \label{fig:app_long_distance}
\end{figure}

In our extended-range experiment for UGreen MAX2 and PHILIPS TAH2020, we increase the injection power from the default 18~dBm to 40~dBm using an external power amplifier (G41P40S) purchased from Alibaba at a cost of \$415. With this higher-power setup, \alias{} can recover audible speech information at distances up to 30~m, extending the attack range beyond the default 6~m setting. The long-distance setup with the power amplifier is shown in~\cref{fig:app_long_distance}. We provide the long-range demos on our project website~\cite{injecteave2025demo}.

%% file: sections/7-Discussion.tex
\section{Discussion}
\label{sec:discussion}

The scope of this work is to provide the theoretical framework and exemplary designs of injection-induced EM side channels, laying the foundation for the scientific characterization of these emerging threats. Based on this framework, this section discusses the need for follow-up research, including exploring other susceptible interfaces and effective protections. 

\subsection{Limitations and Future Work}
\label{sec:discussion on analog input devices}

\textbf{Attack Surface Generalizability.} While our evaluation focused on common audio output devices and smart home applications as motivating and representative examples that carry low-frequency analog secrets, the fundamental leakage models prompt us to extrapolate that \alias{} represents a more generalized attack surface. Theoretically, the vulnerability of injection-induced EM side channels is inherent to any electronic system featuring unshielded nonlinear analog interfaces. We believe future work can contribute to this emerging research topic by examining the susceptibility of a broader class of targets, such as electrical communication signals internal to computer systems, and even confidential data of analog sensor inputs. Below, we discuss our preliminary analysis of audio signals of microphone inputs.

\textbf{Investigation of Analog Microphone Inputs.} 
Our exploration of analog input devices was motivated by an observation in the closed-loop attack on the landline phone in~\cref{sec:casestudy3}: we found that it was also feasible to reconstruct audio from the landline's microphone that captures local user's speech. Then, we further analyzed \alias{} attack on two additional commercial microphones: the UGreen CM769 and the Razer SEIREN V3 MINI.  While we can successfully reconstruct intelligible audio from both devices, our experiment revealed a significantly constrained effective distance compared to the range achieved with headphone audio output. Even under ideal conditions where the target microphones were operating at maximum gain, the maximum range for audio eavesdropping is limited to approximately 30~cm. 
Our analysis shows that this performance disparity is rooted in the amplitude difference between the audio input and output signals. While headphone drivers typically require 1--2~V to operate, microphones generate extremely weak 1--10~mV signals. According to~\cref{eq:Vmod}, this 40--60~dB voltage deficit results in a substantially weaker leakage signal, causing the secret information to be easily obscured by the noise floor even with injection-induced leakage. 

Nevertheless, this current boundary is determined by our EM injection and receiving hardware rather than an intrinsic physical limit. The theoretical maximum eavesdropping distance is $d_{max} \propto \sqrt{|V_{inj}| \cdot |V_{sec}| / N_{sys}}$, as illustrated in~\cref{sec: longer distance eavesdropping}, where $N_{sys}$ denotes the aggregate system noise floor. This shows that the range constraint can be mitigated through further hardware upgrades, such as utilizing high-gain antennas to boost $|V_{inj}|$ and using low-noise spectrum analyzers to suppress $N_{sys}$. We believe this can be achieved by collaborating with resourceful RF researchers as a future work.

\begin{figure}[t]
    \centering
    \includegraphics[width=.9\linewidth]{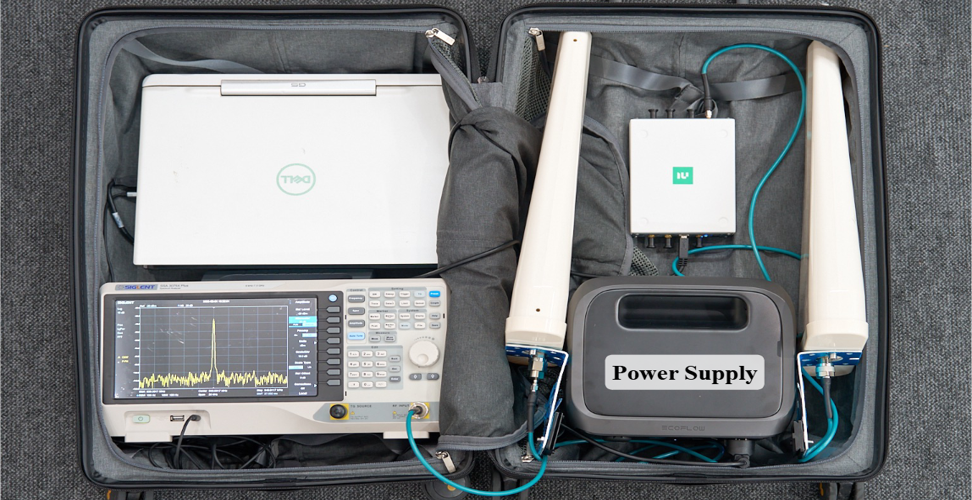}
     \caption{The portable prototype is integrated into a suitcase. Key components include an antenna array, an SDR transceiver, a spectrum analyzer, a laptop, and a portable power supply.}
    \label{fig:prototype}
\end{figure}

\subsection{Mitigation}
We provide insights into potential hardware and software mitigations gleaned from our investigations.

\textbf{Protective Coding Against EM Leakage.}
Existing defenses for mitigating conventional EM side-channel eavesdropping attacks, such as cryptographic masking~\cite{camurati2018screaming} and randomized clocking \cite{onishi2025sound,long2024eye}, are specifically tailored to obfuscate digital logic transitions. Thus, they are mostly ineffective against \alias{} as it eavesdrops on signals in the analog data interfaces. Unlike digital data, these continuous analog waveforms cannot be mathematically masked or randomized without irreversibly degrading the signal fidelity and functional integrity of the device. As a result, we believe more in-depth research is needed to investigate the design space of analog data protection and evaluate utility-security tradeoffs in mitigating emerging analog side-channel leakage.

\begin{figure}
    \centering
     \includegraphics[width=0.99\linewidth]{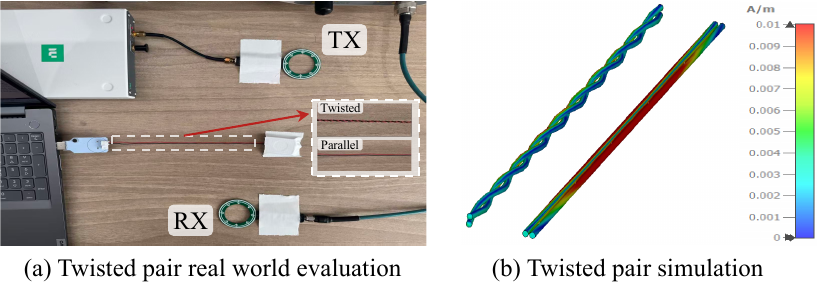}
     \caption{{\color{revcol}Evaluation of twisted-pair mitigation in real-world experiments and simulation.}}
    \label{fig:twisted pair}
\end{figure}

\textbf{Physical Hardening Against EM Injection.}
Traditional EM injection defense typically relies on electromagnetic compatibility (EMC) hardening, such as Faraday shielding~\cite{jiang2023glitchhiker,jin2024phantomlidar}, low-pass filtering~\cite{kune2013ghost}, and differential signaling~\cite{adi_mt095_emi,wang2022ghosttouch}. While these methods can attenuate coupling, they do not guarantee absolute security. According to~\cref{sec: longer distance eavesdropping}, a well-resourced adversary can successfully conduct attacks by increasing the injection power. Furthermore, standard low-pass filters often degrade at high frequencies due to parasitic inductance~\cite{szakaly2024assault}, and specialized high-end filters are rarely viable for consumer electronics.
Consequently, a robust defense paradigm shifts toward active detection and mitigation. Promising approaches include monitoring for anomalous RF carriers~\cite{adami2011hpm} or DC offsets induced by nonlinear rectification~\cite{xu2021inaudible}, alongside software-level consistency checks via sensor fusion~\cite{tu2021transduction} and encoding~\cite{zhang2020detection}.
However, implementing these protections on commercial systems requires addressing significant overheads in manufacturing costs, power consumption, and form factors.

{\color{revcol}
\textbf{Twisted Pair.} We disassemble several devices in Table~\ref{tab: Attacks on COTS Devices} to understand why some devices exhibit shorter eavesdropping ranges. Our teardown analysis reveals that devices using twisted-pair cables exhibited significantly greater resistance to the attack than those with standard parallel wires. To verify this, we build a minimal proof-of-concept setup in~\cref{fig:twisted pair}~(a), evaluating the same USB speaker at 900~MHz with either a standard parallel or a twisted-pair connection, while keeping all other parameters unchanged. The speaker continuously plays a 2~kHz single-tone signal. Under the parallel-pair configuration, the measured injection-induced leakage SNR is 37.6~dB, which decreases to 26.9~dB after switching to twisted-pair wiring, corresponding to a 10.7~dB reduction. To further quantify this effect under ideal conditions, we perform a CST full-wave simulation on a parallel pair with 1.2~mm spacing and an equivalent twisted-pair geometry under identical 900~MHz linearly polarized plane-wave excitation, with the electric field aligned along the wire axis. As shown in~\cref{fig:twisted pair}~(b), the twisted pair exhibits up to 20~dB lower induced surface current than the parallel pair. 
It is worth noting that twisted-pair wiring represents one potentially dominant yet not exclusive contributing factor to leakage susceptibility, as PCB traces, power lines and circuit-level nonlinear components also play a role. While it can serve as a useful starting point for mitigation, achieving full immunity requires a comprehensive, security-aware hardware-software co-design approach from the ground up, and we hope this work can inspire future designers to consider such threats early in the design process. 
}

\section{Related Work}

\textbf{EM Injection and Side-Channel Leakage.}  
Electromagnetic (EM) security has traditionally evolved along two parallel, non-overlapping paradigms: EM injection and side-channel leakage.
EM injection focuses on integrity or availability, where adversaries inject carefully crafted EM signals to induce faults, manipulate sensor output, or disrupt system execution. Recent systematic analysis has shown that EM injection attacks manipulate with multiple stages of cyber-physical system operations, including sensing, computation, actuation, and data communication~\cite{jiang2026sok}, including critical infrastructure~\cite{tu2019trick,yang2024rethink,cyr2023position}, autonomous driving~\cite{jiang2023glitchhiker,jin2024phantomlidar,ren2025ghostshot,zhang2024understanding}, medical healthcare~\cite{kune2013ghost, long2021protecting}, IoT devices~\cite{zhang2024virtual,shan2022invisible,jiang2024ghosttype} and cryptographic modules~\cite{hayashi2012transient,hayashi2014precisely,nishiyama2023remote,menu2020experimental}.
Conversely, EM side-channel leakage focuses on integrity or confidentiality, exploiting unintentional EM emanations to recover the cryptographic key of a target device~\cite{gnad2019leaky,camurati2018screaming,genkin2015stealing} or to reconstruct screen content~\cite{van1985electromagnetic,kuhn2002optical,kuhn2005electromagnetic}, keystrokes~\cite{jin2021periscope}, smartphone displays~\cite{liu2020screen}, biometric data~\cite{li2025emiris,ni2023recovering,xu2025empalm}, and even confidential video streams from a smart home camera~\cite{long2024eye}. 

Existing research largely treats EM injection and EM side-channel leakage as distinct and independent threat vectors, focusing on fault- and disruption-oriented integrity violations and passive confidentiality threats, respectively. 
\alias{} bridges this gap by theoretically and experimentally showing that EM injections can be used to induce side-channel leakage via hardware nonlinearity, redefining the boundary between active integrity injection attacks and passive confidentiality attacks and enabling closed-loop control capabilities of victim devices. This new attack vector offers significantly enhanced capabilities and generalizability compared to existing state-of-the-art eavesdropping methods. For example, while previous passive EM side channels like MagEar~\cite{liao2022magear} and Periscope~\cite{chen2024eavesdropping} are restricted to 0.5--1.5~m, \alias{} can successfully eavesdrop over 
{\color{revcol}30~m. Although DeHiREC~\cite{zhou2023dehirec} uses EM injection to augment the strength of weak EMR signals, it primarily amplifies \textit{existing} side-channel leakage (using $f_3$ to amplify $f_1$) from ADCs to detect whether a hidden voice recorder is ON/OFF. In contrast, InjectEave exploits hardware nonlinearity to induce \textit{entirely new and controllable} side-channel leakage by piggybacking target analog signals $f_1$ onto an injected carrier $f_2$. This different mechanism enables high-fidelity waveform information recovery of continuous analog secrets across diverse electronic components beyond ADCs.} 
Moreover, \alias{} offers a substantially broader attack surface; whereas existing studies are often constrained to specific modalities like audio, our approach exploits fundamental vulnerabilities within the analog systems. This makes \alias{} modality-agnostic and applicable to a vast spectrum of analog devices.

\textbf{Contactless Sensing and Backscattering.}
Contactless sensing and eavesdropping techniques transmit physical signals such as mmWave~\cite{wang2022mmphone,wang2022mmeve,hu2022milliear,basak2022mmspy,li2020wavespy,xu2019waveear, sun2022sok}, acoustic~\cite{cheng2020sonarsnoop,halevi2015keyboard,roy2016listening, ren2023echoimage}, or {\color{revcol}optical}~\cite{sami2020spying,luo2025laserkey,nassi2023little,walker2022laser} probing signals, and analyze the resulting reflections to infer information about a target system. However, most of these contactless sensing technologies are so far only able to sense spatial displacements and movements of objects, such as vocal gesture changes~\cite{li2020wavespy} and human typing activities~\cite{jin2021periscope}. {\color{revcol} 
These prior works rely on the existence of nearby vibrating objects. 
Moreover, optical-based acoustic eavesdropping methods require a clear line of sight, and therefore cannot achieve cross-wall eavesdropping. 
In contrast, \alias{} leverages EM injection and leakage to directly eavesdrop on electrical signals within target circuits, eliminating the dependence on observable physical vibrations and making the attack substantially more robust to environmental movements. Furthermore, the valid EM carrier frequencies, typically in the hundreds-of-MHz range, enable robust through-wall eavesdropping across diverse indoor and outdoor settings.}

The closest line of research to \alias{} is backscattering communications~\cite{yang2024rf,pu2025your,kaji2023echo}, typically used in low-power RFID systems where a transponder intentionally modulates an RF carrier via impedance switching. 
Recent research has extended this to unintentional backscattering. 
{\color{revcol} Several recent works also exploit binary impedance-modulated backscattering to recover serial communication data~\cite{pu2025your,kaji2023echo}, or cryptographic side-channel information~\cite{monfared2023leakyohm,kitazawa2026active}. However, they rely on a preliminary impedance-variation model that limits these works to binary digital data recovery. Analog signals require fundamentally different theoretical modeling and signal interpretation. In contrast, our work builds upon their insights and limitations, and introduces the first eavesdropping-oriented nonlinear model that enables unprecedented long-distance recovery of analog signals including low-frequency secrets on the order of 10~Hz, exposing an orthogonal, scientifically distinct threat surface, which further enables closed-loop EM eavesdropping and manipulation.}

%% file: sections/8-Conclusion.tex
\section{Conclusion}
This work bridges the theoretical and methodological gap between 
conventional side-channel analysis and active electromagnetic injection, introducing the novel threat model and analysis framework of Injection-Induced EM Side Channels. By exploiting the ubiquitous nonlinearity in commodity hardware, we demonstrated that adversaries can actively modulate low-frequency analog secrets onto injected carriers before they are efficiently emitted. 
Our evaluation of 11 commercial devices reveals that this vulnerability is widespread. Our findings underscore an urgent need to develop effective solutions to protect the ubiquitous, but often overlooked, bottom-layer analog interfaces in modern computer systems. 

\textbf{Acknowledgment:}
This work was supported in part by Guangdong Provincial Key Lab of Integrated Communication, Sensing, and Computation for Ubiquitous Internet of Things (No. 2023B1212010007).

%% file: sections/9-ethical-considerations.tex
\section*{Ethical Considerations}

We take strict measures to ensure the safety, legality, and ethical compliance of our research.

\textbf{Stakeholders and Potential Impact.} 
\alias{} has broad implications across multiple stakeholders. Hardware manufacturers of analog audio interfaces, IoT actuators, and power converters may face pressure to adopt differential signaling or enhanced shielding in future designs. 
End users face potential privacy risks regarding home activities and private conversations, highlighting the need for greater public awareness of EM security. Finally, \alias{} advances the EM security research community by providing a theoretical framework for injection-induced leakage, motivating the development of active detection systems and robust defenses.

\textbf{Impact of the Research Process and Publication.} All experiments were performed on electronic devices fully controlled by the authors. To strictly protect privacy, we utilized text-to-speech (TTS) tools to synthesize conversation segments for experimentation rather than recording private conversations of human subjects. The ``through-wall'' and ``hotel room'' scenarios were conducted in cleared and controlled settings, ensuring no third-party systems or non-consenting individuals were targeted. We believe publishing these results is valuable for manufacturers and defenders seeking to understand the limits of EM side-channel eavesdropping and to investigate possible protections. 

\textbf{Mitigation of Negative Impacts.} We have reported these findings to the relevant manufacturers; however, as we have not yet received a response, we have decided to withhold the vulnerable injection frequencies in \cref{tab: Attacks on COTS Devices} and exclude active injection control logics from our open science artifact. We decided to release this security-sensitive information only to trusted and reputable researchers upon request, restricting our research's results to scientific exploration. Furthermore, we prioritize defensive insights, demonstrating that twisted-pair wiring serves as a practical countermeasure by effectively suppressing the induced surface currents.

\textbf{Decision to Conduct and Publish.}
The historical assumption that low-frequency signals are safe from EM leakage has created a false sense of security. We argue that systematically analyzing this emerging phenomenon of injection-induced EM side channel is essential for building effective defenses. Withholding these findings would leave manufacturers blind to the risks. Therefore, we decided to publish this work while removing detailed attack parameters such as vulnerable EM frequencies, and constraining the majority of the paper to theoretical and defensive insights. 

\section*{Open Science}
Demo videos and audio recordings in case studies are available on our website: \url{https://injecteave.github.io/}. Our research artifacts, including received audio recordings in case studies and speech-enhancement codebase are available at: \url{https://doi.org/10.5281/zenodo.20432240}